\documentclass[3p]{elsarticle}
\biboptions{sort&compress}
\usepackage{amsmath}
\usepackage{amsfonts}
\usepackage{amsthm}
\usepackage{graphicx}
\usepackage{graphics}
\usepackage{color}
\usepackage{bbold}
\newcommand{\idmat}{\mbox{$1 \hspace{-1.3 mm} 1$}}
\newcommand{\R}{\mathbb{R}}

\begin{document}

\title{A dispersion and norm preserving finite difference scheme with transparent boundary conditions for the Dirac equation in (1+1)D}
\author[KFU-Graz]{Ren\'{e} Hammer\corref{cor1}}
\ead{rene.hammer@uni-graz.at}
\author[KFU-Graz]{Walter P\"{o}tz}
\ead{walter.poetz@uni-graz.at}
\author[TU-Wien]{Anton Arnold}
\ead{anton.arnold@tuwien.ac.at}
\address[KFU-Graz]{Institut f\"{u}r Physik, Karl-Franzens-Universit\"{a}t Graz, Universit\"{a}tsplatz 5, 8010 Graz, Austria}
\address[TU-Wien]{Institut f\"{u}r Analysis und Scientific Computing, TU-Wien, Wiedner Hauptstr. 8, 1040 Wien, Austria}
\cortext[cor1]{Corresponding author}

\begin{abstract}
 A finite difference scheme is presented for the Dirac equation in (1+1)D.  It can handle space- and time-dependent mass and potential terms and utilizes exact discrete transparent boundary conditions (DTBCs).  Based on a space- and time-staggered leap-frog scheme it avoids fermion doubling and  preserves the  dispersion relation of the continuum problem for mass zero (Weyl equation) exactly. \\
 Considering boundary regions, each with a constant mass and potential term, the associated DTBCs are derived by first applying this finite difference scheme and then using the Z-transform in the discrete time variable. The resulting constant coefficient difference equation in space can be solved exactly on each of the two semi-infinite exterior domains.  Admitting only solutions in $l_2$ which vanish at infinity is equivalent to imposing outgoing boundary conditions.  An inverse Z-transformation leads to exact DTBCs in form of a convolution in discrete time which suppress spurious reflections at the boundaries and enforce  stability of the whole space-time scheme.\\
An exactly preserved functional for the norm of the Dirac spinor on the staggered grid is presented.  Simulations of Gaussian wave packets, leaving the computational domain without reflection, demonstrate the quality of the DTBCs numerically, 
%???
as well as the importance of a faithful representation of the energy-momentum dispersion relation on a grid.
%???
\end{abstract}
\maketitle
\section{Introduction}
We start out with a brief summary regarding important properties of the Dirac equation, its role played in physics, existing numerical schemes for its solution, and the issue of open boundaries.
\subsection{The Dirac equation}\label{intro-dirac}
Next to its fundamental role in relativistic quantum mechanics and field theory, which provide the foundation of  modern nuclear and high energy physics \cite{greiner,thaller}, the Dirac equation has received a rapidly growing importance in condensed matter systems as well. Especially, in the context of the recent experimental realization of graphene \cite{neto}, 2D and 3D topological insulators \cite{hughes,qi}, and optical lattices \cite{lamata,witthaut,szpak} the Dirac equation describes the underlying physics as an effective field theory. Historically it was proposed by Dirac with his ingenious idea of linearizing the square root of the relativistic energy momentum relation by the introduction of Dirac matrices and multi-component wave functions, nowadays known as Dirac spinors. Imposing the condition that the twofold application of this Dirac operator onto the spinor must yield the Klein-Gordon equation, leads to the Clifford algebra for the Dirac matrices. In (1+1)D and (2+1)D the minimum dimension for 
a representation of this group is two, whereas in (3+1)D the Dirac spinor must have a minimum of four components. However, one can work with two components if one accepts higher--order derivatives in space and time \cite{greiner,thaller}.  The latter has lead to the prediction of anti-matter and the concept of the filled Fermi sea in many-particle physics.  Alternatively, the Dirac equation emerges from an investigation of the transformation properties of spinors under the Lorentz group \cite{ryder}. In condensed matter physics Dirac-like equations arise in context of low-energy two-band effective models, e.g. in $k \cdot p$ perturbation theory or the tight-binding approximations \cite{qi}.   Indeed, the study of Dirac fermion realizations has developed into one of the most exciting current topics of condensed matter physics \cite{qi}. 

In this work we restrict ourselves to a study of the (1+1)D  Dirac equation and present a numerically stable scheme for its solution under open boundary conditions.   The latter are motivated by a particle transport situation, as well as the fact that, unlike for the Schr\"{o}dinger particle,  a deep one-particle potential does not ensure confinement for Dirac particles.
In its Schr\"odinger form, also called the standard or Pauli-Dirac form,  the (1+1)D Dirac equation may be written as (using $c=1,\; e=1, \mbox{ and } \hbar=1$) 
\begin{equation}
i \partial_t \mbox{\boldmath$\psi$}(x,t)=\hat{H} \mbox{\boldmath$\psi$}(x,t)\;, \quad\hat{H}= m(x,t) \sigma_z - i \partial_x \sigma_x  -  V(x,t) \idmat_2\;.\label{dirac-eq} 
\end{equation}
The $\sigma_i$'s are the $2\times2$ Hermitian anti-commuting Pauli-matrices, and $\idmat_2$ is the identity matrix. $x\in\mathbb{R},\; t\in\mathbb{R}^{+}$, and 
$\mbox{\boldmath$\psi$}=\left(\begin{array}{c} u \\v\end{array}\right) \in\mathbb{C}^2$ is the complex 2-spinor. $m,V\in\mathbb{R}$ represent, respectively,  a  space- and time-dependent mass and scalar potential. For constant coefficients, Fourier-transformation in the space and time variable and solving the eigenvalue problem  gives the energy spectrum $E_\pm = \pm \sqrt{m^2+p^2}$. In the physical ground state all negative energy states are filled with fermions.  Empty negative energy states are reinterpreted as filled hole states (anti-particles)  with positive energy, in analogy to multi-band systems in (non-relativistic) condensed matter physics \cite{greiner}. The norm of the spinor is defined as $|\mbox{\boldmath$\psi$}(x,t)|_2=\sqrt{|u(x,t)|^2+|v(x,t)|^2}$, and its square can be interpreted as the probability density in space for given time $t$. The norm is conserved because Eq. \
eqref{dirac-eq} is of Schr\"odinger form and $\hat{H}$ is Hermitian \cite{greiner}.  Global gauge-invariance holds as for the case of  the Schr\"odinger equation:  the addition of a constant $V$ simply adds a phase to the solution. The free Dirac equation has various other symmetries \cite{greiner,thallervisual}. First there are the continuous transformations of spatial rotation (only meaningful for more than one space dimension) which, with the Lorentz boost, form the Lorentz group. Together with space and time translations, they form the Poincar\'e group. The latter is the fundamental group in particle physics, but is of minor importance in solid state physics because the crystal lattice necessarily breaks these continuous symmetries. The discrete symmetries holding for arbitrary constants $m\in\mathbb{R}, V\in\mathbb{R}$ are space reflection (parity) and time reversal symmetry which will also be present in the finite difference scheme.

\subsection{Numerical aspects}
Several schemes have been proposed and used for particle transport simulations based on the  time-dependent Dirac equation.
For a numerical treatment one has to discretize the continuum problem either in real- or Fourier-space, or a combination thereof. Real-space schemes are, for example, the  finite-difference \cite{busic, stacey, tworzydlo} and finite-element methods \cite{muller}, whereas the spectral methods \cite{momberger} are examples for the momentum space approach. Split-operator methods separate the time-evolution operator into several parts, with each of them depending only on either momentum or position \cite{braun, mocken}. There also exists a coordinate space split-operator method which transforms the Dirac equation into an advection equation and uses its characteristic solutions  \cite{gourdeau}. While having the advantage of a natural implementation of space- and time dependent potential and mass terms, the finite-difference and finite-element schemes have to deal with the issue of fermion doubling, which means that, for a given sign of the energy,  there are two (or more) extrema in the $m\neq 0$ energy-momentum 
dispersion relation instead of the single one of the continuum problem \cite{stacey}.  In fact the 'Nielsen-Ninomiya no-go theorem' forbids the existence of a single fermion flavor for chirally invariant fermions on a regular grid without breaking either translational invariance, locality, or Hermiticity \cite{nielsen}. In (1+1)D one can get rid of the fermion doubling using a staggered grid for the two spinor components. This is equivalent to taking the left-sided first-order derivative operator for one component of the spinor and the right-sided for the other one \cite{stacey}. One obtains a monotonic dispersion relation with only one minimum. Here we present a scheme which provides  an even better result by applying staggering to both the space and the time coordinate. This yields a numerical scheme which preserves the exact dispersion relation of the continuum problem for the special choice of the ratio between time and space grid $r:=\Delta t/\Delta x = 1$ and mass $m=0$. For $m,V\neq0$ and $r=1$ the 
dispersion relation improves for all possible wave-numbers $k\in[-\pi/\Delta x, \pi/\Delta x]$ with the refinement of the grid. For $\Delta x \rightarrow 0$ the numerical dispersion relation becomes identical to the continuum one. This is not true for most finite difference schemes in general and, to our knowledge, no finite difference scheme with this property for the Dirac equation has been reported before.

Let us mention that numerical methods for the (1+1)D non-linear Dirac (NLD)
equation also got some attention in the literature \cite{xu}. Besides being an
interesting playground on its own, it might also have some physical relevance by
incorporating electron self-interaction into the Dirac equation. Scalar
self-interaction leads to the Soler model \cite{soler}, whereas vector-like
inclusion of the self-interaction leads to the Thirring model \cite{thirring}.
Interestingly, the latter is S-dual to the quantum sine-Gordon model
\cite{thirring}. In contrast to the linear Dirac equation it provides solitary
wave solutions, standing wave solutions, and collapse after collision of two
solitary waves \cite{xu}.
As already for the Dirac equation, analytic solutions are rare, most of the
behavior of the NLD equation can only be investigated numerically \cite{xu}.
For this purpose plenty of algorithms can be found in the literature. Without
claim to completeness, they are of Crank-Nicholson type
\cite{alvarez1,alvarez2,alvarez3}, explicit finite difference schemes
\cite{gordon,xu}, spectral schemes \cite{frutos,wang}, Runge-Kutta methods
\cite{hong,shao}, and moving mesh methods \cite{hwang}.

Remarkably, for the finite difference method no effort was made so far to
eliminate the spurious solutions, e.g.~by staggering the grid for the spinor
components. Being well aware of the different behavior of the NLD equation, we
think our scheme could also have some relevance there. At least it could serve
to solve the linear part in the operator-splitting method, thus avoiding the
transformation to Fourier space.
%???
From a physics perspective, however, the most natural way for the incorporation of self-consistency (e.g., self-interaction) is to solve the standard Dirac equation self-consistently and in parallel to the differential equations for the external potentials which appear in the former.  A brief discussion will be given in the Conclusions in Sect. \ref{conclusions}.  %????
%???
%
%
%
%
\subsection{Boundary conditions\label{sub:General-discussion}}
For a numerical treatment of a differential equation, such as the Dirac equation,  the number of degrees of freedom must be finite.
In addition to the discretization of the time and space variable in a real-space scheme one has to restrict the simulation domain to a finite region in time and space. Then, appropriate boundary conditions are needed to ensure that the solution obtained within the finite domain is (at least)  a good approximation to the solution of the whole space problem. Generally, the time-dependent Dirac equation is solved as an initial-value problem in time.  The standard approach for the derivation of spatial  transparent boundary conditions (TBCs), e.g. for the Schr\"odinger equation,  has been to solve the continuous exterior problem by using the Laplace-transformation in time and to discretize the continuous TBCs afterwards \cite{baskakov}. To avoid stability problems \cite{mayfield} and spurious reflections at the boundary from inconsistent discretization schemes, recently, a new improved approach in which the whole domain is discretized first and then solved exactly in the constant-coefficient exterior domains 
using a Z-transform in time has been developed for the Schr\"{o}dinger equation \cite{arnold, zisowsky}.  In this way one maintains the stability properties of the scheme on the whole space and avoids inconsistent discretization for the simulation region and the boundary conditions. The resulting discrete transparent boundary conditions (DTBCs) which are non-local in time are exact in the sense that they do not introduce a procedural error.  An alternative approach which, in most cases is easier to handle because an inverse Z-transform can be avoided, is to discretize in space only,  derive TBCs, and discretize in time thereafter. This has been done with good results for hyperbolic systems \cite{alonso, lubich, rowley}.  However for the proposed scheme, where the good properties regarding dispersion and conservation of norm arise from the {\it simultaneous} discretization of space and time in an interlaced manner, paying the extra price in form of an inverse Z-transformation is well justified. Therefore we 
follow the fully discrete approach of \cite{arnold} to develop TBCs for the time- and space-staggered leap-frog scheme.  Furthermore, in order to preserve the covariant symmetry of the Dirac equation on a space-time grid, time and space coordinates must be treated on an equal footing.  This is particularly important in higher dimensions.

As a major difference to the Schr\"{o}dinger equation one should mention that a massless relativistic particle cannot be trapped by a scalar one-particle potential \cite{greiner}.  The same holds for massive relativistic particles with energy $E$, when the potential depth $V$ is such that  $E+mc^2> V> E-mc^2$.  This phenomenon is related to the Klein paradox and is due to the two-band-nature of the dispersion relation, consisting of an electron and a positron band. One is necessarily dealing with a scattering problem whenever there are no bound states supported by the Hamiltonian. Then, for initial data which are compactly supported on the computational domain, the solution always reaches the boundary after a finite time.  For simulation times beyond that threshold, open boundary conditions are required  to close the finite difference scheme in a particle transport situation, similar to non-relativistic particle transport simulations in nano devices \cite{talebian,WP-math}.
\\
\section{Continuous transparent boundary conditions for the Dirac equation in (1+1)D\label{1DcontinuousTBCs}}
In this section we derive continuous TBCs for the Dirac equation Eq. \eqref{dirac-eq}. We divide the entire space into the computational domain  $(0,L)$, and the semi-infinite exterior domains $(-\infty, 0]$ and $[L,\infty)$. The mass is assumed to be constant $m(x,t)=m$ and the potential is constant in space $V(x,t)=V(t)$ in the exterior domains. This reduces to the case $V=0$ with the following gauge change of the spinor:
\begin{equation}
\mbox{\boldmath$\psi$}(x,t) = e^{i \mathcal V(t)}\mbox{\boldmath$\chi$}(x,t)\quad\mbox{with}\quad \mathcal V(t) =\int^{t}_{0} V(s) ds~.\label{gauge}
\end{equation}
With $\mbox{\boldmath$\chi$}=(u,v)$, in the exterior, we have
\\
\begin{equation}
i\left(\begin{array}{c}
u(x,t)\\
v(x,t)\end{array}\right)_t=
\left(\begin{array}{cc}
m\quad & -i\partial_x\\
-i\partial_x & -m
\end{array}\right)\left(\begin{array}{c}
u(x,t)\\
v(x,t)
\end{array}\right)~.
\end{equation}\\
The multiplication with $\sigma_x$ and a Laplace transformation with respect to the time variable leads to
\\
\begin{equation}
\left(\begin{array}{c}
\tilde{u}(x,s)\\
\tilde{v}(x,s)\end{array}\right)_x=
\left(\begin{array}{cc}
0\quad & -s + i m\\
-s - i m & 0
\end{array}\right)\left(\begin{array}{c}
\tilde{u}(x,s)\\
\tilde{v}(x,s)
\end{array}\right)~.\label{laplace-transformed}
\end{equation}\\
Eq. \eqref{laplace-transformed} has the general solution
\\
\begin{equation}
\left(\begin{array}{c}
\tilde{u}(x,s)\\
\tilde{v}(x,s)\end{array}\right)=
c_1 \left(\begin{array}{cc}
-s + im\\
-\sqrt[+]{s^2+m^2}
\end{array}\right) e^{-\sqrt[+]{s^2+m^2} x}
+c_2 \left(\begin{array}{cc}
-s + im\\
\sqrt[+]{s^2+m^2}
\end{array}\right) e^{\sqrt[+]{s^2+m^2} x}
~,\label{gen-sol}
\end{equation}
where $\sqrt[+]{~~}$ is written for the square root with positive real part. Because the solution must be in $L_2(\R)$ the constant $c_2$ must vanish on the right exterior domain. For the same reason $c_1$ must be zero on the left exterior domain. Therefore, the boundary conditions on the right boundary are
\begin{equation}
\partial_x \tilde{u}(x,s)\big |_{x=L} = -\sqrt[+]{s^2+m^2}~\tilde{u}(L,s)\qquad\mbox{and}\qquad \partial_x\tilde{v}(x,s)\big |_{x=L} = -\sqrt[+]{s^2+m^2}~\tilde{v}(L,s)~.
\end{equation}
On the left boundary one gets
\begin{equation}
\partial_x \tilde{u}(x,s)\big |_{x=0} = \sqrt[+]{s^2+m^2}~\tilde{u}(0,s)\qquad\mbox{and}\qquad \partial_x\tilde{v}(x,s)\big |_{x=0} = \sqrt[+]{s^2+m^2}~\tilde{v}(0,s)~.
\end{equation}
The structure for both spinor components and both boundaries is the same, so we proceed with $\tilde u$ on the right boundary. First we derive boundary conditions as a Neumann-to-Dirichlet map by writing 
\begin{equation}
\tilde{u}(L,s) = -  \frac{1}{\sqrt[+]{s^2+m^2}} \partial_x \tilde{u}(x,s)\big |_{x=L}~.\label{NtDs}
\end{equation}
Then the inverse Laplace transformation leads to the convolution at the right boundary:
\begin{equation}
u(L,t) = -  J_0(m t) \ast_{t} \partial_x u(x,t)\big |_{x=L}
=-\int_0^t J_0(m\tau)  \partial_x u(L,t-\tau)\,d\tau~,
\end{equation}
with $J_0$ being the Bessel function of first kind. Analogously for the left boundary:
\begin{equation}
u(0,t) =   J_0(m t) \ast_{t} \partial_x u(x,t)\big |_{x=0}~.
\end{equation}\\
Finally, this leads with the gauge Eq. \eqref{gauge} for $V\neq0$ to the TBCs in the form of a Neumann-to-Dirichlet map:
\begin{equation}
\mbox{\boldmath$\psi$}(L,t) = - e^{i \mathcal V_r(t)}\Big\{ J_0(m t) \ast_{t}\Big[ \partial_x \mbox{\boldmath$\psi$}(x,t)\big |_{x=L} e^{-i \mathcal V_r(t)}\Big]\Big\}\quad\ldots\mbox{right TBC}~,\label{NtD-rcTBC}
\end{equation}
\begin{equation}
\mbox{\boldmath$\psi$}(0,t) =  e^{i \mathcal V_l(t)}\Big\{ J_0(m t) \ast_{t} \Big[\partial_x \mbox{\boldmath$\psi$}(x,t)\big |_{x=0} e^{-i \mathcal V_r(t)}\Big]\Big\}\quad\ldots\mbox{left TBC}~.\label{NtD-lcTBC}
\end{equation}\\
\\
For the derivation of TBCs in the form of a Dirichlet-to-Neumann map we write:
\begin{equation}
\partial_x \tilde{u}(x,s)\big |_{x=L} = - \tilde g(s) \tilde{u}(L,s)~,
\end{equation}
with:
\begin{equation}
\tilde g(s):=\sqrt[+]{s^2+m^2} = \Bigg(\frac{s^2+m^2}{\sqrt[+]{s^2+m^2}}-s\Bigg)+s~.
\end{equation}
We use
\begin{equation}
\tilde f(s) := \frac{1}{\sqrt[+]{s^2+m^2}}\quad \stackrel{\mathcal L^{-1}}{\rightarrow}\quad f(t)= J_0(m t)
\end{equation}
and
\begin{equation}
s^2\tilde f(s) - s f(0) \quad \stackrel{\mathcal L^{-1}}{\rightarrow}\quad m^2 J''_0(m t)~.
\end{equation}\\
This leads to the following inverse Laplace transform of $\tilde g(s)$:
\begin{equation}
\tilde g(s)\quad \stackrel{\mathcal L^{-1}}{\rightarrow}\quad g(t) = m^2 \big[J''_0(m t)+J_0(m t)\big] + \delta' = \frac{m^2}{2} \big[J_2(m t)+J_0(m t)\big] + \delta' ~,
\end{equation}
where $\delta'$ is the first derivative of the delta distribution.
Then the TBC as a Dirichlet-to-Neumann map on the right boundary is
\begin{equation}
\partial_x \mbox{\boldmath$\psi$}(x,t) \big |_{x=L} = - e^{i \mathcal V_r(t)}\Big\{\frac{m^2}{2} \big[J_2(m t)+J_0(m t)\big]\ast_{t}\mbox{\boldmath$\psi$}(L,t) e^{-i \mathcal V_r(t)} + \partial_t \mbox{\boldmath$\psi$}(L,t) e^{-i \mathcal V_r(t)}\Big\}  ~.\label{DtN-rcTBC}
\end{equation}
On the left boundary one gets
\begin{equation}
\partial_x \mbox{\boldmath$\psi$}(x,t) \big |_{x=0} = e^{i \mathcal V_r(t)}\Big\{\frac{m^2}{2} \big[J_2(m t)+J_0(m t)\big]\ast_{t}\mbox{\boldmath$\psi$}(0,t) e^{-i \mathcal V_r(t)} + \partial_t \mbox{\boldmath$\psi$}(0,t) e^{-i \mathcal V_r(t)}\Big\}  ~.\label{DtN-lcTBC}
\end{equation}
Discretizations of Eqs. \eqref{NtD-rcTBC} and \eqref{NtD-lcTBC} or Eqs. \eqref{DtN-rcTBC} and \eqref{DtN-lcTBC} can serve as boundary conditions for arbitrary finite difference discretizations of the Dirac equation Eq. \eqref{dirac-eq}. But one has to be aware of the fact that inconsistent discretization of the differential equation and the associated boundary conditions usually leads to spurious reflections or even instability. As already mentioned in the introduction we will therefore first apply the discretization scheme to Eq. \eqref{dirac-eq} for the the boundary regions also and derive the associated TBCs by Z-transformation. Eqs. \eqref{NtD-rcTBC}, \eqref{NtD-lcTBC}, \eqref{DtN-rcTBC}, and \eqref{DtN-lcTBC} serve as a guide to gain intuition for the behavior of the convolution coefficients.
\section{Time- and space-staggered leap-frog scheme}
Leap-frog time-stepping, in combination with a staggered spatial grid, plays a special role among the finite difference methods because, in addition to the elimination of the fermion-doubling problem, it proves to be dispersion relation preserving in (1+1)D for the `golden ratio' of $r=\Delta t/ \Delta x=1$, $m=0$, and $V=0$ (Weyl equation).  Moreover, for $m\neq0$ and/or $V\neq0$ the dispersion relation is still monotone and improves with a refinement of the grid. It is identical to the exact analytic dispersion when $\Delta t, \Delta x\rightarrow 0,$ for fixed $r=1$. This is particularly important for simulations where the whole possible range of the wave numbers $k \in \left[ - \frac{\pi }{\Delta x},\frac{\pi}{\Delta x}\right]$ is used, for example, in strong external fields.  An initial wave-packet which consists only of wave-components near $k=0$  may acquire high wave number components due to strong spatial and/or temporal changes of potential and/or mass. 

For schemes with non-monotonic dispersion a problem can also arise at the boundary. The modes of such a scheme consist of additional, spurious, numerically generated modes on the lattice which, nevertheless, must be accounted for in the DTBCs for consistency. This requires special attention at the boundary because improper realization of the boundary condition may lead to `energy' transfer between the modes, spurious reflections, and eventually instability (see e.g. \cite{rowley}).  A correct dispersion relation means correct phase and group velocity on the grid and is essential for faithful long-time propagation studies.  We now present such a scheme.
\subsection{The discretization scheme}
We shall consider the following leap-frog discretization of the Dirac equation in Pauli-Dirac form given in Eq.\eqref{dirac-eq}:
\begin{align}
\frac{u_{j}^{n+1/2}-u_{j}^{n-1/2}}{\Delta t}+i(m_j^n-V_j^n)\frac{u_{j}^{n+1/2}+u_{j}^{n-1/2}}{2}
+\frac{(D v^n)_j}{\Delta x}&=0\;,\label{eq:discretization-leap-frog-halfgrid1} \\ 
\frac{v_{j-1/2}^{n+1}-v_{j-1/2}^{n}}{\Delta t}-i(m_{j-1/2}^{n+1/2}+V_{j-1/2}^{n+1/2})\frac{v_{j-1/2}^{n+1}+v_{j-1/2}^{n}}{2}
+\frac{(D u^{n+1/2})_{j-1/2}}{\Delta x}&=0\;,
\label{eq:discretization-leap-frog-halfgrid}
\end{align}
\\
with $j\in\mathbb{Z},\; n\in\mathbb{N}$. Here we used the notation $\mbox{\boldmath
$\psi$}(x,t) =  \big[u(x,t),v(x,t)\big]$ with $ u(x_j,t^{n-1/2})\approx u_j^{n-1/2} $ and $ v(x_{j-1/2},t^{n})\approx v_{j-1/2}^{n}$. The symmetric spatial difference operator is defined as $(D v^n)_j = v_{j+1/2}^{n}-v_{j-1/2}^{n}$ and $(D u^{n+1/2})_{j-1/2} = u_{j}^{n+1/2}-u_{j-1}^{n+1/2}$. For the mass and potential term we use the averages $g_j^n=\big(g_{j}^{n+1/2}+g_{j}^{n-1/2}\big)/2$ for the integer spacial grid-points and $g_{j-1/2}^{n+1/2}=\left(g_{j-1/2}^{n+1}+g_{j-1/2}^{n}\right)/2$ for the half-integer ones, where $g=m,V$.
%???
The space-time stencil is shown in Fig. \ref{fig:Staggered-grid, leap-frog} (a). As one can see, the components $u$ and $v$ are not defined on the same space-time grid points but on sub-grids shifted by a half time and space grid spacing. A Taylor expansion about the grid points $(x_{j},t_{n})$ and $(x_{j-1/2},t_{n+1/2})$ shows that the staggered-grid leap-frog scheme is second order accurate in space and time.\\
The leap-frog time stepping is not self starting.  Assume, for example,  initial data $\mbox{\boldmath$\psi$}(x,t)$ given at the time-level $t^0$. Then the $u$ component has to be propagated from $t^{0}$ to $t^{1/2}$. In principle this can be done by any time stepping method. We have chosen a symplectic Euler step \cite{hairer} with half space-time grid spacing, which is algorithmically equivalent to the leap-frog scheme, with the exception that all the data (u and v) are stored at the same time and space level. Moreover it is only first order accurate (omitting indices for $m$ and $V$):
\begin{align}
\frac{u_{j}^{1/2}-u_{j}^{0}}{\Delta t/2}+i(m-V)\frac{u_{j}^{1/2}+u_{j}^{0}}{2}
+\frac{v_{j}^{0}-v_{j-1/2}^{0}}{\Delta x/2}&=0~,\quad j\in \mathbb{Z}/2.
\end{align}
After this first initialization step the $u$-component is stored only at integer and the $v$-component at half integer spatial points. With a relabeling of the indices, the algorithm Eq. \eqref{eq:discretization-leap-frog-halfgrid} can be written as follows:
\begin{align}
\frac{u_{j}^{n+1}-u_{j}^{n}}{\Delta t}+i(m-V)\frac{u_{j}^{n+1}+u_{j}^{n}}{2}
+\frac{v_{j}^{n}-v_{j-1}^{n}}{\Delta x}&=0\;,\nonumber \\ 
\frac{v_{j}^{n+1}-v_{j}^{n}}{\Delta t}-i(m+V)\frac{v_{j}^{n+1}+v_{j}^{n}}{2}
+\frac{u_{j+1}^{n+1}-u_{j}^{n+1}}{\Delta x}&=0~,
\label{eq:discretization-leap-frog}
\end{align}
with $j\in \mathbb{Z}$, $n \in \mathbb{N}_0$. Here we use the approximations $u_j^n\approx u(x_{j-1/4},t^{n-1/4})$,  $v_j^n\approx v(x_{j+1/4},t^{n+1/4})$ (see Fig. \ref{fig:Staggered-grid, leap-frog} (b)).
\begin{figure}[t!]
\centering
\includegraphics[width=9cm]{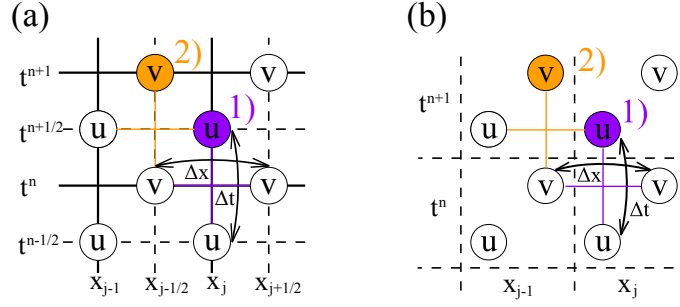}
\caption{(color online). (a) Staggered-grid scheme for Dirac equation in Pauli-Dirac
form with leap-frog time-stepping; (b) the algorithmically equivalent (except for starting procedure) symplectic Euler time-stepping. \label{fig:Staggered-grid, leap-frog}}
\end{figure}
Rearranging of terms leads to the following equations for the explicit recursive update
\begin{align}
u_{j}^{n+1} &= \frac{2-i(m-V) \Delta t}{2+i(m-V) \Delta t} u_{j}^{n} - \frac{2 \Delta t/\Delta x}{2+i(m-V) \Delta t} \Big(v_{j}^{n}-v_{j-1}^{n}\Big)\;,\\
v_{j}^{n+1} &= \frac{2+i(m+V) \Delta t}{2-i(m+V) \Delta t} v_{j}^{n} - \frac{2 \Delta t/\Delta x}{2-i(m+V) \Delta t} \Big(u_{j+1}^{n+1}-u_{j}^{n+1}\Big)\;.\label{expl-form}
\end{align}
The starting procedure for the leap-frog scheme with a symplectic Euler step of half grid-size is especially suitable since symplectic Euler is algorithmically equivalent to the leap-frog staggered-grid scheme and has the same dispersion relation for equal ratio $r$.
\subsection{Von Neumann stability analysis}
For constant coefficients in Eq. \eqref{eq:discretization-leap-frog}, Fourier analysis can be used to perform a von Neumann stability analysis and to derive the dispersion relation for the whole space problem. 
A Fourier transform in space of Eq. \eqref{eq:discretization-leap-frog}  leads to
\begin{equation}
\underbrace{\left(\begin{array}{cc}
\frac{1}{\Delta t}+\frac{i(m-V)}{2} & 0\\
\frac{e^{-ik\Delta x}-1}{\Delta x} & \frac{1}{\Delta t}-\frac{i(m+V)}{2}
\end{array}\right)}_{=:\mathbf{A}}
\left(\begin{array}{c}
\tilde{u}^{n+1}\\
\tilde{v}^{n+1}\end{array}\right)
+
\underbrace{\left(\begin{array}{cc}
-\frac{1}{\Delta t}+\frac{i(m-V)}{2} & \frac{1-e^{ik\Delta x}}{\Delta x}\\
0 & -\frac{1}{\Delta t}-\frac{i(m+V)}{2}
\end{array}\right)}_{=:\mathbf{B}=-\mathbf{A}^\ast}
\left(\begin{array}{c}
\tilde{u}^{n}\\
\tilde{v}^{n}\end{array}\right)
=0\;.\label{xFT}
\end{equation}\\
Using the definitions $\xi = k \Delta x$, $\mu = m \Delta t$, $\nu = V \Delta t$, and  $r=\Delta t/\Delta x$ the  eigenvalues  of the amplification matrix
\begin{equation}
\mathbf{G} = - \mathbf{A}^{-1} \mathbf{B}=\mathbf{A}^{-1} \mathbf{A}^\ast=\left(\begin{array}{cc}
\frac{2 i - \nu+\mu}{2 i + \nu-\mu} & \frac{2 i r(e^{i \xi}-1)}{2 i + \nu-\mu}\\
\frac{2 r (1-e^{-i \xi})(2 + i \nu-i \mu)}{4-\nu^2 - 4 \nu i +\mu^2} & \frac{4 + \nu^2-\mu^2 + 4 \mu i + 8 r^2 [\cos \xi-1]}{4-\nu^2 - 4 \nu i +\mu^2}
\end{array}\right)
\end{equation}
are computed by means of
\begin{equation}
\lambda_\pm=  P/2 \pm \sqrt{\big(P/2\big)^2-Q}~,
\label{growthfactor}
\end{equation}
where $P = \mbox{tr} [\mathbf{G}]$ and $Q = \mbox{det}[\mathbf{G}]$, with
\begin{equation}
P = \big[2(\nu^2-\mu^2)+8(1-r^2+r^2\cos \xi)\big]/N\label{P}
\end{equation}
and
\begin{equation}
Q = \big[\mu^2-(\nu-2 i)^2\big] /N\label{Q}~,
\end{equation}
where $N=\mu^2-(\nu+2 i)^2$. A lengthy but straightforward computation yields $|\lambda_{\pm}|=1$ for $r\leq1$. The eigenvalues are non-degenerate except for the case $k=\mu=0$ and $r=1$ (with geometric multiplicity 2), as well as the case $k=\pi/\Delta x$, $\nu=0$, and $r=1$ (with geometric multiplicity 1). Except for the latter case, the scheme is stable under the constraint $r\leq1$, which constitutes its Courant-Friedrichs-Lewy (CFL) condition. 
Below we use a different technique to prove stability which allows one also to identify a functional (related to the $l_2$-norm defined on the staggered grid) which is exactly conserved.
\\
\begin{figure}[t!]
\centering
\includegraphics[width=16.45cm]{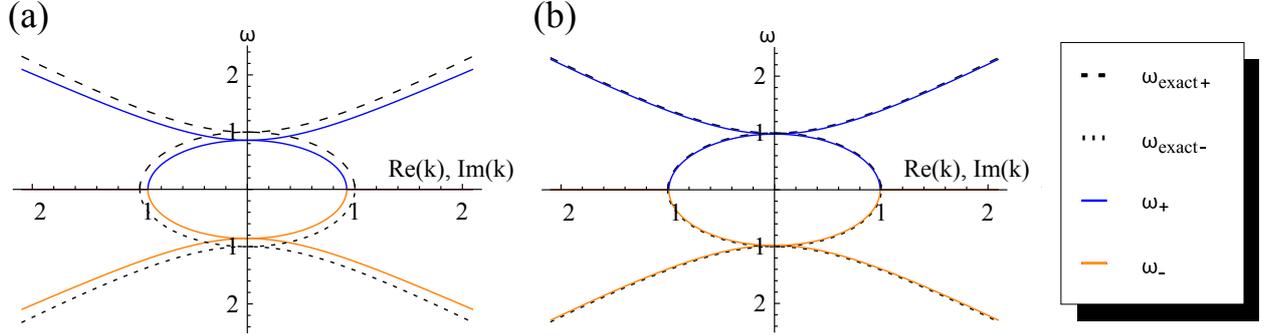}
\caption{(color online). Numerical dispersion relation $\omega(k)$ for the
staggered-grid leap-frog scheme with $m=1$ and $r=1$ for real $k$
(hyperbolically shaped curves) and for imaginary $k$ inside the energy gap
(elliptically shaped curve) shown as solid lines. The dispersion relation for
the continuum problem is shown for comparison using dotted lines: (a) very
coarse grid $\Delta x = 1.5$ and $k\in[\pi/\Delta x,\pi/\Delta x]$; (b) finer
grid $\Delta x = 0.5$ where merely an excerpt of the wave number range
$k\in[\pi/ (3 \Delta x),\pi/(3 \Delta x)]$ is shown.\label{coarsefine}}
\end{figure}
\subsection{The energy-momentum dispersion relation}
\begin{figure}[t!]
\centering
\includegraphics[width=6cm]{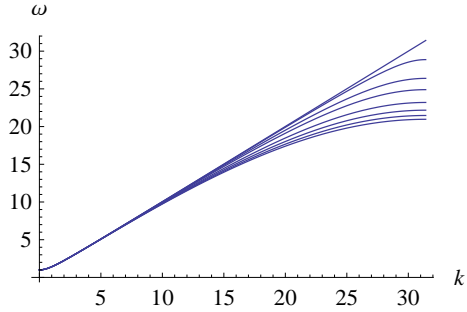}
\caption{Dispersion relation $\omega(k)$ for $V=0$, $m=1$, and  $\Delta x = 0.1$, and $k\in[0,\pi/\Delta x]$ with $r=1,0.99,0.95,0.9,0.8,0.7,0.6,0.5$ (from top to bottom).}\label{disp_rvar}
\end{figure}
\begin{figure}[t!]
\centering
\includegraphics[width=10cm]{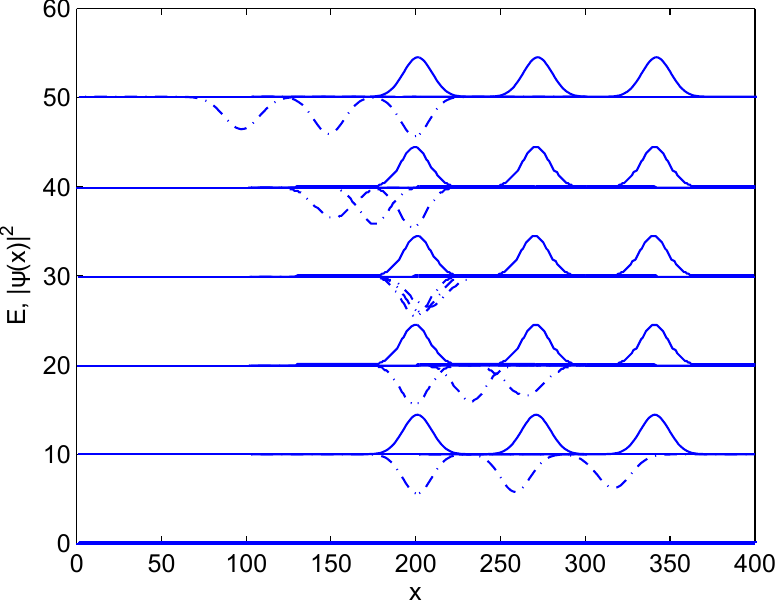}
\caption{Comparison of the leap-frog staggered-grid scheme with a scheme
that is centered in space and Crank-Nicolson in time. We have choosen
$\Delta t = \Delta x = 0.05$ and $m=V=0$. The Gaussian wave packet is initially
placed at the center ($x=200$) and should then move to the right. We give the
results for 5 different energy mean values ($E=10,..., 50$). For each
prescribed energy we merge the simulation results at three times ($t=0,70,140$) into
one graphics. The figures show the probability density
$|\mbox{\boldmath$\psi$}(x)|^2$ (solid lines for the leap-frog
staggered-grid scheme) and the negative probability density (dash-dotted lines)
for the Crank-Nicolson scheme. The latter shows large dispersive errors, whereas
the leap-frog staggered-grid scheme shows the correct
propagation.}\label{fermiondoubling}
\end{figure}

The Fourier transform of Eq. \eqref{xFT} with respect to the time variable with  $\tilde{\omega}=\omega \Delta t$ leads to the homogeneous system
\begin{equation}
(e^{i \tilde{\omega}} \mathbf{A} + \mathbf{B}) \mbox{\boldmath$\tilde{\psi}$} = 0~,
\end{equation}
with the solutions for $r\leq 1$:
\begin{equation}
\tilde{\omega}_\pm = -\frac{i}{r} \ln [\lambda_{\pm}] ~\label{grid-dis},
\end{equation} 
where $\lambda_{\pm}$  is given in Eq. \eqref{growthfactor}.
By setting  $\mu,\nu=0$ they reduce to
\begin{equation}
\tilde{\omega}_\pm = - \frac{i}{r} \ln \Bigg\{1 + r^2 \big(\cos\xi-1\big)\pm \sqrt{\Big[ r^2 \big(\cos\xi-1\big)+1\Big]^2-1} \Bigg\},
\end{equation}  
\noindent and with the choice $r=1$ they read
\begin{equation}
\tilde{\omega}_\pm = - i \ln \Big(\cos\xi \pm \sqrt{\cos^2\xi-1} \Big) = -i \ln \Big(\cos \xi \pm i \sin \xi \Big) = \pm \xi~.\label{perfect}
\end{equation}
Thus for $\mu,\nu=0$, and $r=1$ the linear dispersion of the continuum problem (Weyl equation) is exactly preserved.\\
\\
The connection to the phase of the growth factor (i.e. eigenvalues of $\mathbf{G}$)  can be established via
\begin{equation}
\tilde{\omega}_\pm = -\frac{i}{r} \ln [\lambda_{\pm}] = -\frac{i}{r} [\ln |\lambda_{\pm}| + i \arg (\lambda_{\pm}) + 2 \pi i n] = \frac{1}{r} [\arg (\lambda_{\pm}) + 2 \pi n] ~.
\end{equation}
In Fig. \ref{coarsefine} the dispersion relation for a rather coarse grid with $\Delta x=1.5$ is compared to that of a finer one using $\Delta x=0.5$,  for  $m=1$, $V=0$, and $r=1$.  Clearly, the quality of the numerical dispersion relation improves for a finer grid and, for all wave-numbers $k\in[-\pi/\Delta x,\pi/\Delta x]$, it approaches the continuum form in the limit  $\Delta x \rightarrow 0$ and  $r=1$. In Fig. \ref{disp_rvar} the dispersion relation for fixed values of $m$ and $V$ is shown for several values of $r$ and $\Delta x=0.1$.\\
\\
In Fig. \ref{fermiondoubling}, the scheme is compared to a scheme using a
centered approximation for the space derivative and Crank-Nicolson time
averaging, showing the fermion-doubling problem. The comparison is made for
$m=0=V$, $r=1$, and initial Gaussian wave packets with different mean energies.
The Crank-Nicolson scheme shows large dispersive errors and for high mean energy
(large wavenumbers) and even propagation in the wrong direction
(fermion-doubling) occurs, whereas the leap-frog staggered-grid scheme yields
(almost) the exact solution.
\\
\\
\subsection{Phase error and gauge invariance}
\begin{figure}[t!]
\centering
\includegraphics[width=16.45cm]{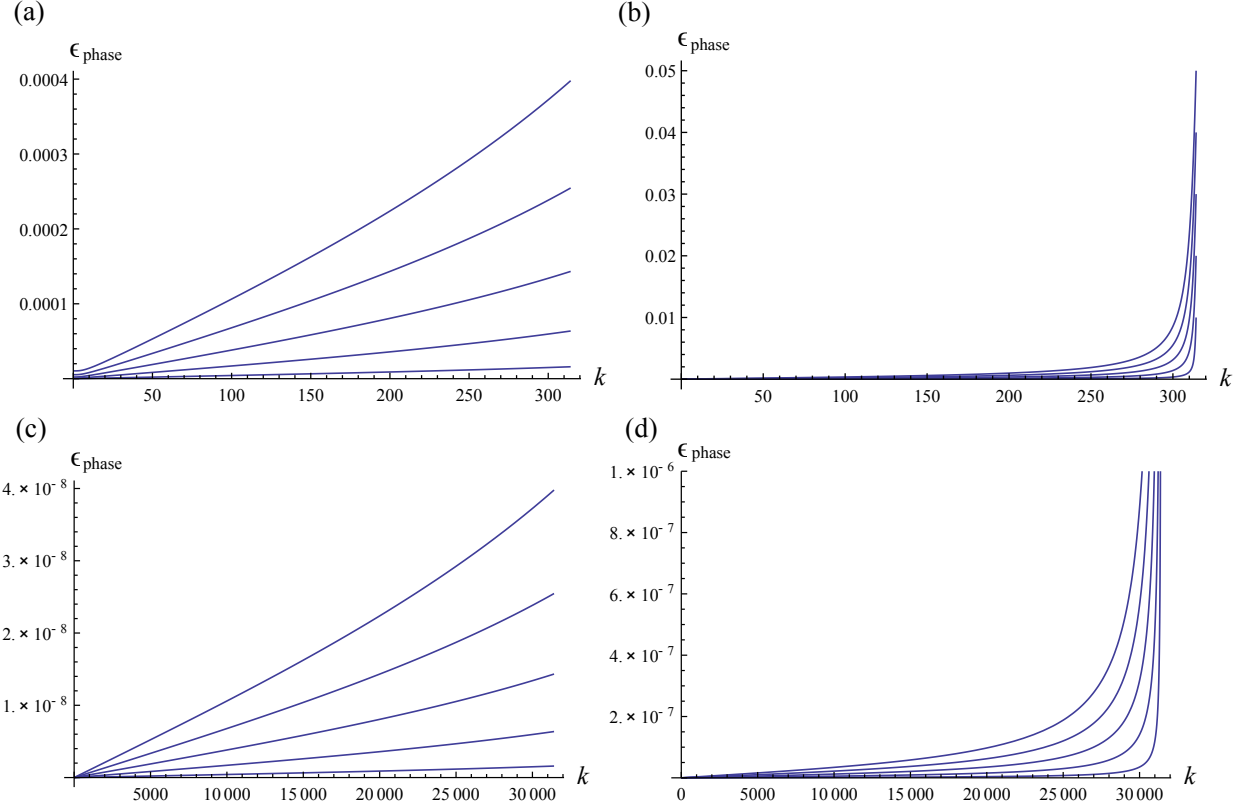}
\caption{The dependence of the phase error $\epsilon_{\mbox{\scriptsize phase}}$ on the wave vector $k\in[0,\pi/\Delta x]$ for $V=0$, $m=1,2,3,4,5$ (from bottom to top): (a) coarse grid $\Delta t =\Delta x = 0.01$, (c) fine grid $\Delta t =\Delta x = 10^{-4}$; for $V=1,2,3,4,5$ (from bottom to top), $m=0$: (b) coarse grid $\Delta t =\Delta x = 0.01$, (d) fine grid $\Delta t =\Delta x = 10^{-4}$. In (d) the functions exceed the plot-range, where the maxima for $V=(1,2,3,4,5)$ are $\epsilon_{\mbox{\scriptsize phase}} = (1,2,3,4,5)*10^{-4}$ at $k=10^{4} \pi$. \label{phaseerror}}
\end{figure}

One can define the phase error for one time-step as
\begin{equation} 
\epsilon_{\mbox{\scriptsize phase}}(k,r,\Delta t)=\big[\omega_{\mbox{\scriptsize analytical}}(k)-\omega_{\mbox{\scriptsize numerical}}(k,r,\Delta t)\big]\Delta t~,
\end{equation}
where the analytic dispersion relation is $\omega_{\mbox{\scriptsize analytical}}=\pm\sqrt{m^2+k^2}$ and the numerical one is given in Eq. \eqref{grid-dis}. Together with Eq. \eqref{perfect} it follows that $\epsilon_{\mbox{\scriptsize phase}}$ vanishes in the case of $m=V=0$ and $r=1$ where the scheme indeed propagates the solution without error. In Fig. \ref{phaseerror}, $\epsilon_{\mbox{\scriptsize phase}}$ is shown for different values of $m$ and $V$ on a coarse grid $\Delta t = \Delta x = 0.01$ (a) and (b), and on a finer grid $\Delta t = \Delta x = 10^{-4}$ (c) and (d).\\
\\
Due to gauge invariance, when adding a diagonal term $V$ (constant scalar potential) to the Hamitionan Eq. \eqref{dirac-eq} one can introduce a new spinor 
\begin{equation}
\breve{\psi}(t) = \psi(t) \exp \big(- i V t)~,
\end{equation}
which fulfills the original equation. Therefore the gauge error per time-step is equivalent to $\epsilon_{\mbox{\scriptsize phase}}$ due to a finite $V$ (see Fig. \ref{phaseerror} (b) and (d)). The gauge invariant introduction of the electromagnetic vector potential is shown in \chapter{B}.
\subsection{Stability analysis within a multiplication technique} 
The von Neumann analysis above and the fact that the dispersion relation $\omega(k)$ is real for $r\neq 1$ revealed the  stability conditions of the scheme for constant mass and potential terms. 
However, there is another technique available which is not based on the  Fourier transform and leads to the identification of an ``energy" functional which is exactly conserved by the scheme, even in the presence of non-constant  mass and potential terms. The use of this multiplication technique has been inspired by a  stability analysis of a leap-frog pseudo-spectral scheme  for the Schr\"{o}dinger equation \cite{borzi}.\\

We define the inner product $(u,v) := \sum_j u_j \bar{v}_j$ on $l^2(\mathbb{Z};\mathbb{C})$ and we will use the notation $\left\|u\right\|^2:=(u,u).$  Taking the inner product of Eq. \eqref{eq:discretization-leap-frog-halfgrid1} with $(u^{n+1/2}+u^{n-1/2})$ and taking the real part gives
\begin{equation}
\left\|u^{n+1/2}\right\|^2-\left\|u^{n-1/2}\right\|^2 + r \Re\Big[(D v^n,u^{n+1/2}+u^{n-1/2})\Big]  = 0~.\label{firsteq}
\end{equation}
Eq. \eqref{eq:discretization-leap-frog-halfgrid} is multiplied by $(\bar{v}^{n+1}+\bar{v}^{n})$ and again the real part is taken to give 
\begin{equation}
\left\|v^{n+1}\right\|^2-\left\|v^{n}\right\|^2 + r \Re\Big[(D u^{n+1/2},v^{n+1}+v^{n})\Big]  = 0~.\label{secondeq}
\end{equation}
Performing a summation by parts with vanishing boundary terms at infinity gives
\begin{equation}
 \Re \Big[(D v^n, u^{n+1/2}+u^{n-1/2})\Big]=-\Re\Big[(D u^{n+1/2}+D u^{n-1/2}, v^n)\Big]~.
\end{equation}
Then adding Eq. \eqref{firsteq} and Eq. \eqref{secondeq} leads to

\begin{equation}
\left\|u^{n+1/2}\right\|^2 + \left\|v^{n+1}\right\|^2 + r \Re \Big[(D u^{n+1/2},v^{n+1})\Big] = \left\|u^{n-1/2}\right\|^2 + \left\|v^{n}\right\|^2 + r \Re \Big[(D u^{n-1/2},v^{n})\Big]
\end{equation}
and one immediately identifies the conserved functional
\begin{equation}
E_r^n := \left\|u^{n+1/2}\right\|^2 + \left\|v^{n+1}\right\|^2 + r \Re \Big[(D u^{n+1/2},v^{n+1})\Big] = \mbox{const}= E_r^0~.\label{E}
\end{equation}
With this result one obtains the stability condition for the scheme by using
\begin{equation}
 \Big|\Re \Big[(D u^{n+1/2},v^{n+1})\Big]\Big|\leq\left\|u^{n+1/2}\right\|^2 + \left\|v^{n+1}\right\|^2~.
\end{equation}
This gives the estimate
\begin{equation}
\left\|u^{n+1/2}\right\|^2 + \left\|v^{n+1}\right\|^2\leq \frac{E_r^0}{1-r} \quad \forall~n~,
\end{equation}
\\for $r<1$.
The case $r=1$ must be treated separately and one rewrites $E_1^n$ as
\begin{align}
E_1^n&=\sum_j |u_j^{n+1/2}|^2 + |v_{j+1/2}^{n+1}|^2 + \Re \sum_j \big(u_{j+1}^{n+1/2}-u_j^{n+1/2}\big)\bar{v}_{j+1/2}^{n+1}\\
&=\frac{1}{2} \sum_j |u_{j}^{n+1/2}-v_{j+1/2}^{n+1}|^2 + \frac{1}{2} \sum_j |u_{j+1}^{n+1/2}+v_{j+1/2}^{n+1}|^2~.\label{t1}
\end{align}
Alternatively, when shifting indices, one obtains
\begin{equation}
E_1^n=\frac{1}{2} \sum_j |u_{j}^{n+1/2}+v_{j-1/2}^{n+1}|^2 + \frac{1}{2} \sum_j |u_{j}^{n+1/2}-v_{j+1/2}^{n+1}|^2~.\label{t2}
\end{equation}
Using $\frac{1}{4}\left\|a_1 + a_2\right\|^2 \leq \frac{1}{4} \big(\left\|a_1 + b\right\|+\left\|a_2 - b\right\|\big)^2\leq \frac{1}{2} \left\|a_1 + b\right\|^2 + \frac{1}{2} \left\|a_2 - b\right\|^2~$ gives
\begin{align}
\mbox{from Eq. \eqref{t1}:\qquad} \left\|\tilde{u}^{n+1/2}\right\|^2:= \sum_j \Bigg|\frac{u_{j}^{n+1/2}+u_{j+1}^{n+1/2}}{2}\Bigg|^2&\leq E_1^n = E_1^0\quad \forall n\\\nonumber
\mbox{from Eq. \eqref{t2}:\qquad\quad} \left\|\tilde{v}^{n+1}\right\|^2:= \sum_j \Bigg|\frac{v_{j-1/2}^{n+1}+v_{j+1/2}^{n+1}}{2}\Bigg|^2~&\leq E_1^n = E_1^0\quad \forall n
\end{align}
\begin{equation}
\Rightarrow\quad \left\|\tilde{u}^{n+1/2}\right\|^2 + \left\|\tilde{v}^{n+1}\right\|^2 \leq 2 E_1^0~.
\end{equation}\\
One easily verifies that $\|\tilde u\|$ is a norm on $l^2(\mathbb{Z})$. Indeed,
$\tilde u=0$ implies $u_j=(-1)^j\lambda$ for some $\lambda\in\mathbb{C}$. And
$u\in l^2$ then yields $u=0$.

This allows us to conclude that the scheme is stable for all $r=\Delta t /\Delta
x\leq 1$. Moreover we have identified the functional which is conserved by the
scheme (see Eq. \eqref{E}). In fact, it is conserved for arbitrary time and
space dependent $m, V \in \mathbb{R}$.\\
\\
\subsection{Time-reversal invariance}
The time reversal invariance of the scheme can easily be seen in Eqs. (20) and  \eqref{eq:discretization-leap-frog-halfgrid}. One has to set $\Delta t \rightarrow -\Delta t$ and replace the role of the old and new time-levels $n-1/2\leftrightarrow n+1/2$ and $n\leftrightarrow n+1$. Then one observes that the scheme for the backward propagation has exactly the same form as for the forward propagation and concludes the scheme is time-reversal invariant.
\section{Discrete transparent boundary conditions for the staggered-grid leap-frog scheme\label{1DDTBCs}}
Having discussed the properties of the leap-frog scheme we now turn to the derivation of the associated TBCs.  Again,  we divide the entire space into the computational domain  $(0,L)$, a left semi-infinite exterior domain $(-\infty, 0]$,  and a right semi-infinite exterior domain $[L,\infty)$.  This corresponds to a typical device simulation geometry (scattering scenario) in which the nano-device is placed in the computational domain and the (macroscopic) contacts
are represented by the exterior domains.
 We make the following simplifying assumptions:
\vphantom{}
\begin{itemize}
\item The initial data $\mbox{\boldmath$\psi$}_0(x)=\mbox{\boldmath$\psi$}(0,x)$ is compactly
supported inside the computational domain.
\item In each exterior domain the mass $m(x,t)=m$ and potential $V(x,t)=V$ are constant in $t$ and \nolinebreak $x$. 
\end{itemize}
Both assumption are made for simplicity but can be loosened when needed, as will be discussed below.\\

We first Z-transform Eq. \eqref{eq:discretization-leap-frog} in $t$-direction on each of the exterior domains, $j \in \{\ldots,-2,-1,0\}$ and $j \in  \{J,J+1,J+2,\ldots\}$,  and solve the resulting finite difference equations explicitly. 
Using the definition of the Z-transform $b(z):=Z(b_{n})=\overset{\infty}{\underset{n=0}{\sum}}b_{n}z^{-n}$, its shifting property
$Z(b_{n+1})=\overset{\infty}{\underset{n=0}{\sum}}b_{n+1}z^{-n}=\overset{\infty}{\underset{n\text{�}=1}{\sum}}b_{n\text{�}}z^{-n\text{�}+1}=z b(z)-z b_{0}$,  and setting $b_{0}=0$ (since the initial spinor is compactly supported on $(0,L)$ ) we obtain
\\
\begin{align}
&\left(\begin{array}{cc}
\frac{1}{\Delta t}(z-1)+\frac{i(m-V)}{2}(z+1) & \quad\frac{1}{\Delta x}\\
\frac{z}{\Delta x} & \quad 0 
\end{array}\right)\left(\begin{array}{c}
u_{j}(z)\\
v_{j}(z)
\end{array}\right)\\\nonumber
&\qquad\qquad+\left(\begin{array}{cc}
 0\quad  & -\frac{1}{\Delta x}\\
-\frac{z}{\Delta x}\quad & \frac{1}{\Delta t}(z-1)-\frac{i(m+V)}{2}(z+1)
\end{array}\right)\left(\begin{array}{c}
u_{j-1}(z)\\
v_{j-1}(z)
\end{array}\right)=0\;.\label{difference-eq-raw}
\end{align}\\
\\
Translation from grid point $j$ to $j+1$ therefore is given by 
\\
\begin{equation}
\left(\begin{array}{c}
u_{j}\\
v_{j}
\end{array}\right)=\left(\begin{array}{cc}
1\quad & \frac{\Delta x}{\Delta t}\frac{1-z}{z}+\frac{i(m+V) \Delta x}{2}  \frac{1+z}{z}\\
 \frac{\Delta x}{\Delta t}(1-z)-\frac{i(m-V) \Delta x}{2} (1+z)\quad & 1+\frac{\Delta x^2}{z}\Big[\frac{(1-z)^2}{\Delta t^2} + i V \frac{1-z^2}{\Delta t} + \frac{(m^2-V^2)(z+1)^2}{4}\Big]
\end{array}\right)\left(\begin{array}{c}
u_{j-1}\\
v_{j-1}
\end{array}\right)~.\label{translation}
\end{equation}\\
This can also be written as
\\
\begin{equation}
\left(\begin{array}{c}
u_{j}-u_{j-1}\\
v_{j}-v_{j-1}
\end{array}\right)=\underbrace{\left(\begin{array}{cc}
0\quad & \frac{\Delta x}{\Delta t}\frac{1-z}{z}+\frac{i(m+V) \Delta x}{2}  \frac{1+z}{z}\\
 \frac{\Delta x}{\Delta t}(1-z)-\frac{i(m-V) \Delta x}{2} (1+z)\quad & \frac{\Delta x^2}{z}\Big[\frac{(1-z)^2}{\Delta t^2} + i V \frac{1-z^2}{\Delta t} + \frac{(m^2-V^2)(z+1)^2}{4}\Big]
\end{array}\right)}_{=:\mathbf{M}}\left(\begin{array}{c}
u_{j-1}\\
v_{j-1}
\end{array}\right)~,\label{M}
\end{equation}\\
where $\mathbf{M}:=\left(\begin{array}{cc}
0 & a\\
b & c
\end{array}\right)$ fulfills $a b=c$.
Solving the system Eq. \eqref{M} leads to
\begin{equation}
u_{j+1} - (2+c) u_j + u_{j-1} = 0~,
\end{equation}
whose characteristic equation has the roots:
\begin{equation}\label{eq54}
\tau_{1,2} = 1 + \frac{c}{2} \pm \sqrt{c+\frac{c^2}{4}}.
\end{equation}
The same result is obtained for the component $v$. Since $\tau_1 \tau_2 = 1$, there is one decaying mode (as $j\rightarrow \infty$) with $|\tau_1| \leq 1$ and one increasing mode with $|\tau_2| \geq 1$. In order to have a solution in $l_2(\mathbb{Z})$ one has to choose the mode with $|\tau_1| \leq 1$.  At each boundary only one spinor component couples to the contact region.  The spinor components $u$  and $v$, respectively, only need a right  and a left TBC (see Eq. \eqref{eq:discretization-leap-frog} and Fig. \ref{fig:Staggered-grid, leap-frog}) 
\begin{align}
v_0(z) &= \tau_1(z) v_1(z)\\\nonumber
u_J(z) &= \tau_1(z) u_{J-1}(z)
\end{align}
Then, the inverse Z-transformed boundary conditions are in the form of a convolution in the discrete time variable:
\begin{equation}
v_{0}^{n}=\overset{n}{\underset{k=0}{\sum}}\tau_1^{(n-k)}v_{1}^{k}~.\label{convolution}
\end{equation}
\begin{equation}
u_{J}^{n}=\overset{n}{\underset{k=0}{\sum}}\tau_1^{(n-k)}u_{J-1}^{k}~.
\end{equation}
\begin{figure}[t!]
\centering
\includegraphics[width=11cm]{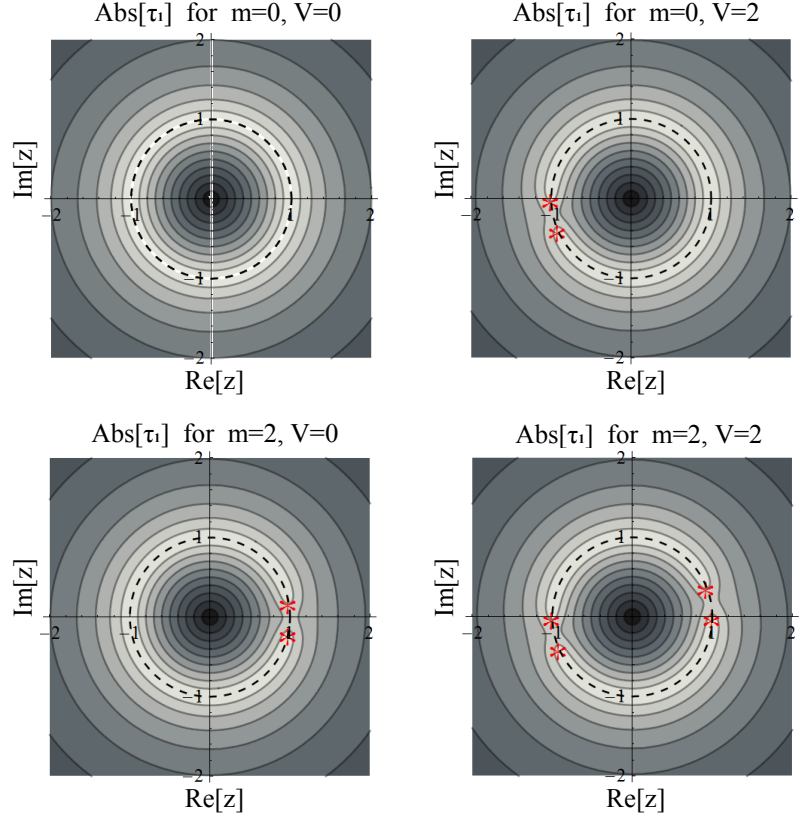}
\caption{Absolute value of $\tau_1=1/\tau_2$ which is zero at $|z|=0$ and $|z|=\infty$, shown for $\Delta t = \Delta x =0.1$. The contour lines are at 0.1, 0.2, 0.3 $\ldots$ etc. The branch-point locations are marked by (red) stars. \label{branchpoints}}
\end{figure}

$\tau_{1}(z)$ is a non-rational function of $z$, hence there is no easy way of finding an analytic expression for its inverse Z-transformed in general. However, the poles and branch-points in the $z$-plane, which determine the general behavior of their inverse Z-transformed in (discrete) real time $n$, can be identified.  We give a brief discussion of these special points for $\Delta:=\Delta t = \Delta x$.\\
First, one observes that $\tau_1$ has no poles. For the branch cuts we examine the branch points of the square root function -- due to zeros of its argument.
This leads to the four branch points of $\tau_1$ (Fig. \ref{branchpoints}):
\begin{equation}
z_{1} =-1~,\quad z_{2} = -\frac{4 (1+i V \Delta)+(m^2-V^2)\Delta^2}{4 (1-i V \Delta)+(m^2-V^2)\Delta^2} \quad\mbox{and}\quad z_{3,4} = \frac{2 i - (V\pm m)\Delta}{2 i + (V\pm m)\Delta}~.\label{z34}
\end{equation}\\
It can easily be seen that they all lie on the unit circle. These branch points induce damped oscillations in time
for the convolution coefficients $\tau_1^{(n)}$.  The high frequency part $(-1)^n$ of these oscillations, which arises from $z_{1} =-1$,  can introduce numerical problems because of subtractive cancellation. It may be advantageous to eliminate this behavior:  A multiplication of $\tau_1$(z) by $(z+1)/z$ allows one to construct new coefficients as a linear combination of the original coefficients involving the two time steps $\tilde{\tau}^{(k)} = \tau^{(k)}+\tau^{(k-1)}$  (see \cite{arnold, antoine}).
Then \eqref{convolution} becomes
$$
  v_{0}^{n}=\overset{n-1}{\underset{k=0}{\sum}}\tilde\tau_1^{(n-k)}v_{1}^{k}-v_0^{n-1}~,
$$
where we used $\tau_1^0=\lim_{z\to\infty} \tau_1(z)=0$.

\begin{figure}[t!]
\centering
\includegraphics[width=16.45cm]{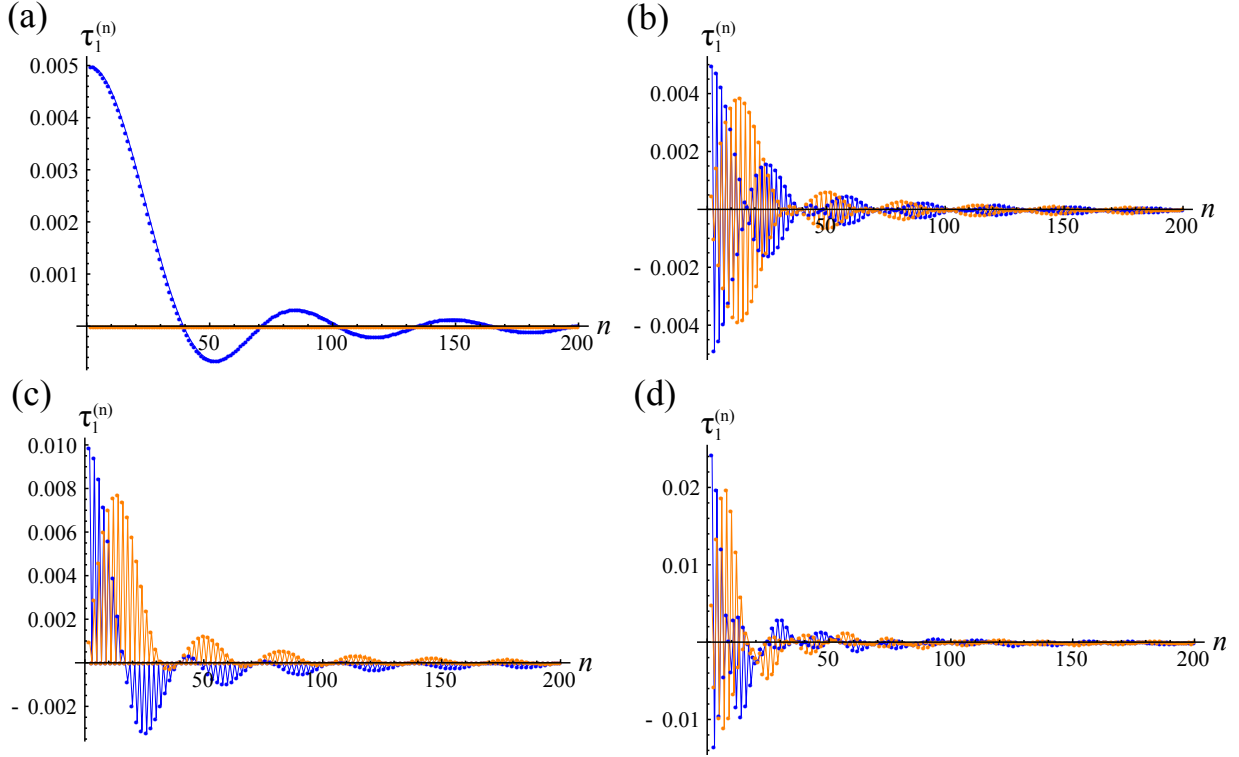}
\caption{(color online). Decay of the convolution coefficients $\tau_1^{(n)}$ for different values of $m$ and $V$  for $\Delta t = \Delta x = 0.1$: (a) $m=1$ and $V=0$, (b) $m=0$ and $V=1$, (c) $m=1$ and $V=1$, (d) $m=1$ and $V=2$. The darker (blue) color shows the real part of the coefficients, the brighter (orange) color shows the imaginary part. The zeroth coefficient $\tau_1^{(0)}$ is always zero and is not shown in the plots. The first coefficient exceeds the plot-range and has the values: (a) $\tau_1^{(1)}=0.9975$, (b) $\tau_1^{(1)}=0.9925+0.0995 i$, (c) $\tau_1^{(1)}=0.9901+0.0990 i$, (d) $\tau_1^{(1)}=0.9682+0.1951 i$~. }\label{coeff}
\end{figure}

For our numerical simulations, the inverse Z-transformation was carried out by performing a power series expansion about $z=0$ using Mathematica. In Fig. \ref{coeff} we show the convolution coefficients $\tau_1^{(n)}$ for $\Delta t= \Delta x=0.1$ and various values of $m$ and $V$.
For the special case $m=V=0$ and $r=1$ the inverse Z-transform can be computed analytically. One gets 
\begin{equation}
\tau_1 = \frac{1}{2}\Big(z+\frac{1}{z}\Big) - %\pm 
\frac{1}{2}\sqrt{\Big(z-\frac{1}{z}\Big)^2} =\ldots = \frac{1}{z}~.
\end{equation}
%where again $\pm$ indicates that the correct branch of the square root has to be chosen. 
The convolution coefficient  then simply  is  $\tau_1^{(n)}=\delta_1^n$, where $\delta_k^n$ is the Kronecker symbol. In the \ref{A} we shall give an explicit, analytic derivation of the convolution coefficients $\tau_1^{(n)}$ in the general case.
\subsection{Numerical examples}

With the general properties of the scheme established and the associated TBCs identified, we now put it to test in challenging numerical applications. 

\begin{figure}[t!]
\centering
\includegraphics[width=16cm]{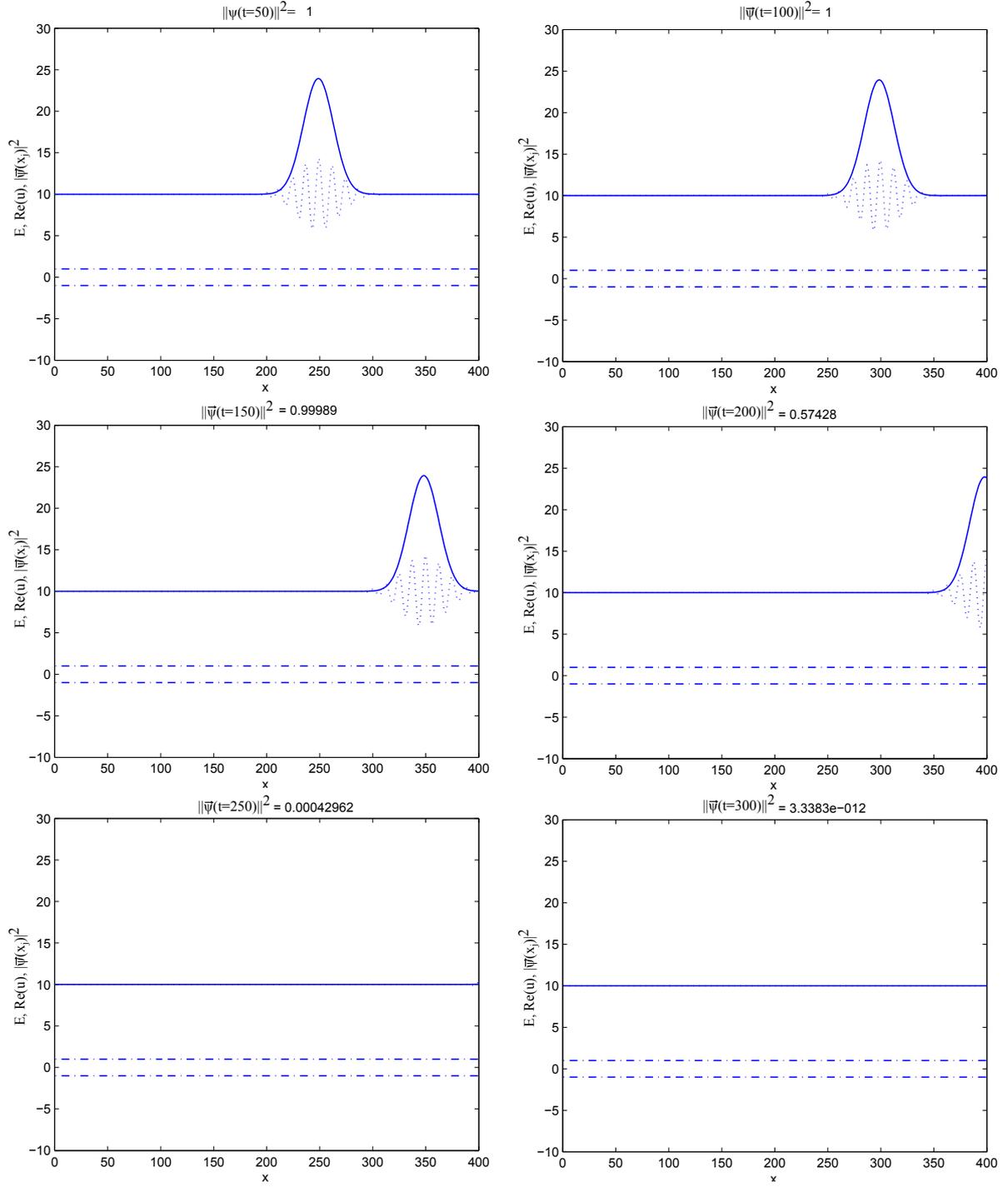}
\caption{Simulation run for a coarse grid $\Delta t = \Delta x= 0.05$ and an initial Gaussian wave packet with a mean value of $k = 15.92 \%$ of $k_{max}$ and a standard deviation of $\sigma_k = 1.59 \%$ of $k_{max}$. The figures show the probability density $|\mbox{\boldmath$\psi$}(x)|^2$ (solid lines), the real part of the upper component $u$ (dotted lines), and the mass-gap (point-dotted lines). The zero line on the vertical axis is shifted by $10$, which is the mean energy of the wave packet. With  time evolving, the probability $\|\mbox{\boldmath$\psi$}\|^2$  for finding the particle in the computational domain approaches zero because the wave packet leaves the domain without reflection.}\label{sim1}
\end{figure}

In Fig. \ref{sim1}  we show  the results of a simulation run for a Gaussian wave packet starting out with a mean value of $k = 15.92 \%$ of $k_{max}$ and a standard deviation of $\sigma_k = 1.59 \%$ of $k_{max}$. A coarse grid $\Delta t = \Delta x= 0.05$ is used. The individual figures show the probability density $|\mbox{\boldmath$\psi$}(x,t)|^2$ computed with
\begin{equation}
|\psi_j^n|^2 :=  |u_j^{n+1/2}|^2 + |v_{j+1/2}^{n+1}|^2 + \Re \Big[ \big(u_{j+1}^{n+1/2}-u_j^{n+1/2}\big)\bar{v}_{j+1/2}^{n+1}\Big]~,\label{densitywithconservedE}
\end{equation}
which gives the conserved functional Eq. \eqref{E} when summed over $j$ on an infinite grid. Additionally, the real part of the upper component $u$, and the mass-gap (= energy region between the minimum of the electron band and the maximum of the positron/hole band) is shown. One can observe that the wave packet leaves the simulation domain without reflections (the numerical error is  less than machine precision = 2.2204e-16).  

\begin{figure}[t!]
\centering
\includegraphics[width=16.45cm]{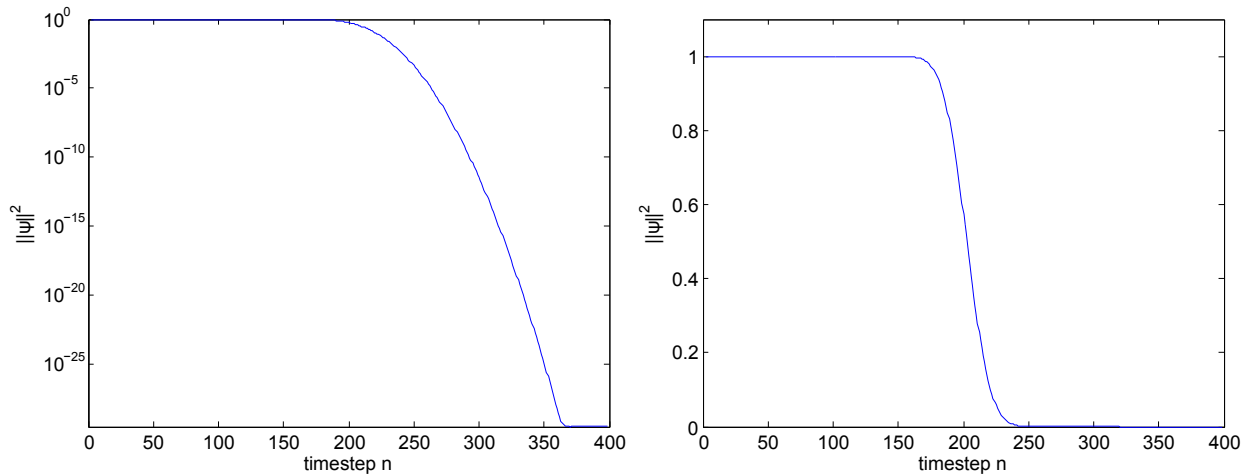}
\caption{Time dependence of the spinor norm $\|\mbox{\boldmath$\psi$}(t)\|$ of the simulation run shown in Fig. \ref{sim1}: logarithmic scale (left) and linear scale (right).}\label{norm1}
\end{figure}

In Fig. \ref{norm1} the time dependence of the probability for finding the Dirac particle inside the computational interval $\|\mbox{\boldmath$\psi$}(t)\|^2$ (:= Eq. \eqref{densitywithconservedE} summed over $j$ on the computational domain) is shown for this simulation.

\begin{figure}[t!]
\centering
\includegraphics[width=15.9cm]{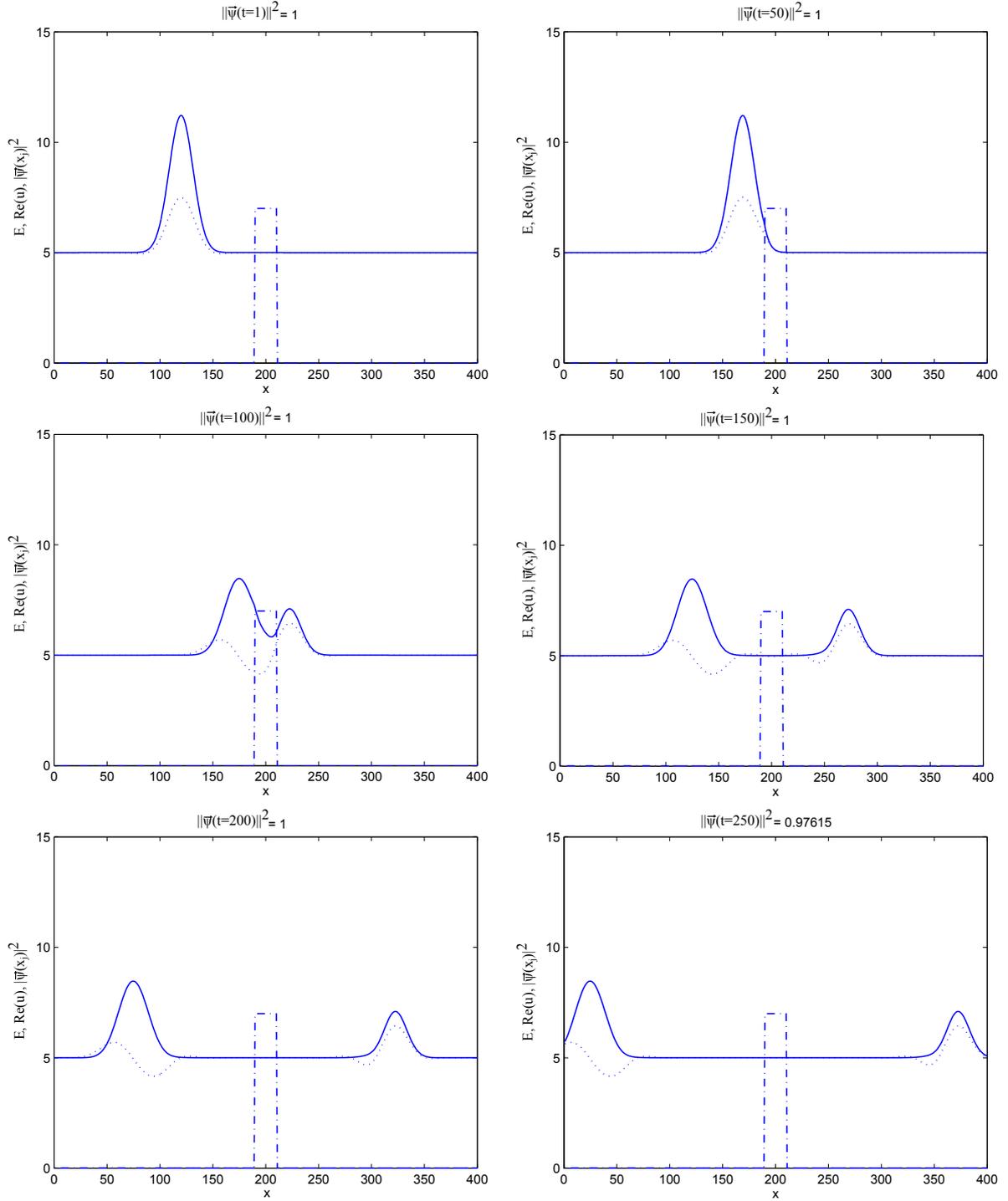}
\caption{Reflection at a mass barrier with a width of 25 grid-points and a height of $m=7$. $\Delta t = \Delta x= 0.01$, the initial Gaussian wave packet has a mean value of $k = 1.59 \%$ of $k_{max}$ and a standard deviation of $\sigma_k = 1.99 \%$ of $k_{max}$. The individual figures show the probability density $|\mbox{\boldmath$\psi$}(x)|^2$ (solid lines), the real part of the upper component $u$ (dotted lines), and the mass (dash-dotted lines). The zero for the vertical axis is shifted by $5$, which corresponds to the mean energy of the initial wave packet.}\label{sim2}
\end{figure}

In Fig. \ref{sim2} we show the reflection at a mass barrier (= spatial region where the energy of the wave packet lies inside the mass-gap) with $m=7$ and a width of 25 grid-points placed near the center of the simulation region. A variation of the mass term can be achieved for Dirac fermions on topological insulators by magnetic texturing \cite{hammerAPL,hammerDW}. The wave packet initially has a mean value of $k = 1.59 \%$ of $k_{max}$ and a standard deviation of $\sigma_k = 1.99 \%$ of $k_{max}$, where $\Delta t = \Delta x= 0.01$. In this tunneling problem it is of profound importance that the dispersion relation is correct also for imaginary $k$, which is the case for the proposed scheme already for relatively coarse grids, as illustrated in  Fig. \ref{coarsefine}.   

\begin{figure}[t!]
\centering
\includegraphics[width=16cm]{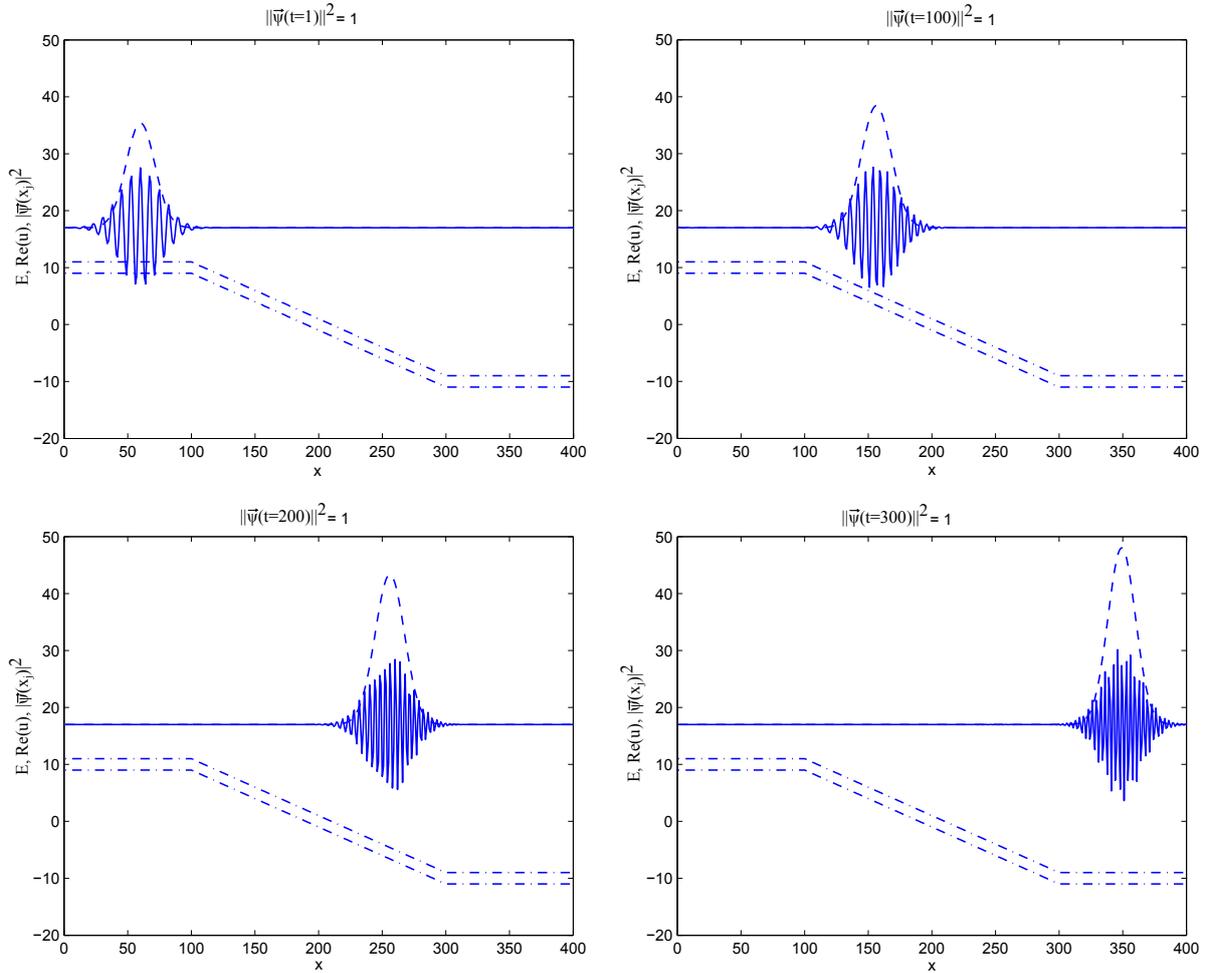}
\caption{Initial Gaussian wave packet with a mean value of $k = 27.06 \%$ of $k_{max}$ and a standart deviation of $\sigma_k = 1.99  \%$ of $k_{max}$ meeting a linear potential. A rather coarse grid $\Delta t = \Delta x= 0.05$ is used. The figures show the probability density $|\mbox{\boldmath$\psi$}(x)|^2$ as dashed line, the real part of the upper component $u$ as solid line, and the mass-gap as dash-dotted lines. The zero line is shifted by $17$ which is the value for the mean energy of the wave packet.}\label{sim3}
\end{figure}

Fig. \ref{sim3} shows a simulation for an initial wave packet with a mean value of $k = 27.06 \%$ of $k_{max}$ and a standard deviation of $\sigma_k = 1.99  \%$ of $k_{max}$ moving across a linear potential drop. In this process, the wave number grows beyond the maximum wave number provided by the grid $k_{max}$, where the wave packet is under-sampled.  The propagation is still correctly described because the dispersion and therefore the group velocity is well approximated for all possible wave numbers resolved by the grid.\\
\subsection{Non-compactly supported initial condition for the spinor  and time-dependent exterior potential}
\noindent The constraints stated at the beginning of chapter \ref{1DDTBCs} can be loosened:
\begin{itemize}
\item The DTBCs can readily be generalized to the case where the initial wave packet is not compactly supported inside the computational domain $x\in(0,L)$. The inhomogeneous boundary conditions can be derived by substituting $\mbox{\boldmath$\psi$}(x_{r,l},t)$ with $\mbox{\boldmath$\psi$}(x_{r,l},t) - \mbox{\boldmath$\psi$}_{in}(x_{r,l},t)$ and carrying out the procedure as detailed in chapter \ref{1DDTBCs}.  
Here, $\mbox{\boldmath$\psi$} - \mbox{\boldmath$\psi$}_{in}$ should initially have compact support in $(0,L)$. And
$\mbox{\boldmath$\psi$}_{in}$ should be a solution to the discrete exterior domain problem, e.g.\ a \emph{discrete} plain wave (see \cite{Ar01} for details in the analogous Schr\"odinger case).
Re-substitution afterwards leads to the result
\begin{equation}
u_{J}^{n}=\overset{n}{\underset{k=0}{\sum}}\left(\tau_1^{(n-k)}u_{J-1}^{k}\right)+u_{J,in}^{n}\;,
\end{equation}
\begin{equation}
v_{0}^{n}=\overset{n}{\underset{k=0}{\sum}}\left(\tau_1^{(n-k)}v_{1}^{k}\right)+v_{0,in}^{n}\;.
\end{equation}
The rationale is that the Dirac equation is a linear differential equation which here is approximated by a linear finite difference scheme. Therefore one may simply add inhomogeneous boundary terms via the superposition principle.

\item 
The scalar potential $V(t)$ on the exterior domain may depend on time. This is useful when considering a time-dependent net potential drop across the system.  As for the case of the Schr\"odinger equation a time-dependent exterior potential, e.g. at the right boundary, can be incorporated by treating it in the interaction picture which removes  $V(t)$ from the Hamiltonian and leads to a gauge (phase change) for the spinor Eq. \eqref{gauge} \cite{antoine}.

\end{itemize}
\newpage
\section{Conclusions and future work\label{conclusions}}
%???-begin
In this paper we have presented a finite difference scheme for the numerical solution of the time-dependent Dirac equation in (1+1)D, allowing  fully time- and space-dependent mass and potential terms.  Owing to a combined staggering of the grid both in space and time, unphysical additional Dirac cones are avoided and, for the special case of the Weyl equation,  the linear dispersion is preserved exactly for all wave numbers supported by the grid.  In the case of finite mass and/or potential terms the dispersion relation improves for all wave numbers, approaching the continuum dispersion relation exactly, when the grid is refined. This is a relevant feature when modeling Dirac fermions on a lattice, such as Dirac fermions propagating on topological insulator surfaces, graphen, or in quantum spin Hall states. The electromagnetic potentials accounting for an external electromagnetic field are included in gauge invariant fashion.  A stability analysis of the scheme was performed and a functional,  exactly conserved by the scheme,  was identified.  It provides a valid norm for the spinor on the proposed staggered grid.  

Furthermore, we have derived exact DTBCs to close the finite difference scheme for constant mass and potential in the boundary regions. With these BCs one is in the position to deal with particle transport scenarios and to account for the multi-band nature of the Dirac equation which may cause inter-band transfer rather than quantum confinement.  For completeness,  DTBCs were also derived for the (1+1) Dirac (differential) equation in Schr\"{o}dinger form.
 
Using the norm for a measure, numerical simulations of Gaussian wave packets show them leaving the computational domain without reflections (with an error below computer precision), thus verifying the quality of the DTBCs numerically.  The importance of a faithful representation of the energy-momentum dispersion relation, in particular, avoiding fermion doubling, is exemplified in numerical simulations.  

The assumptions of constant mass and potential on the exterior domain can be loosened. It is however desirable that an analytic solution on the discrete exterior domain can be found. The case of purely time-dependent exterior potential can be treated by using a proper phase change of the spinor. Initial conditions for which the initial wave packet is not compactly supported on the computational domain can be handled as well, leading to inhomogeneous terms in the boundary conditions. There are various ways to extend the proposed leap-frog scheme to the (2+1)D Dirac equation where it retains many of its attractive features \cite{cpcpaper,ArXiVhammer,singlecone}. The formulation of DTBCs for the (2+1)D Dirac equation in the spirit of this paper is the subject of future work \cite{singlecone}.\\
\\
In view of existing work on  non-linear versions of the Dirac equation accounting, for example, for self-interaction corrections in a single equation, we wish to point out that a self-consistent treatment of such effects can be treated readily (and in more generality) within the present approach based on the standard Dirac equation and a parallel self-consistent update of the effective electromagnetic potentials.  In a semi-classical picture, for example, the latter may be accomplished within Maxwell's theory for the electromagnetic potentials \cite{jackson}.  In this fashion, a self-consistent open-boundary treatment of the Schr\"{o}dinger equation was performed whereby, instead of the nonlinear Hartree term, equivalently the Poisson equation was solved in parallel \cite{talebian}.  Any nonlinearity arising from the (classical) electromagnetic interaction can readily be treated within this approach and its advantages be exploited, as long as the conditions on the asymptotic regions made for the linear Dirac equation can be met.  This is the case, for example, when non-linear effects are confined to the simulation region and/or nonlinear effects in the outer regions can be accounted for by  constant electromagnetic potential and mass terms.  Clearly, the direct use of non-linear versions of the Dirac equation requires the use of individually matched transparent boundary conditions, as for the case of the Schr\"{o}dinger equation \cite{antoine}.  %????
%???_end
\section*{Acknowledgments}
We thank B. A. Stickler and C. Ertler for stimulating discussions.\\
We also gratefully acknowledge funding by the Austrian Science Foundation under project I395-N16.
\\
\\
%
%-------APPENDIX-------
%
\newpage
\begin{appendix}
\section{Analytic derivation of the convolution coefficients $\tau_1^{(n)}$ for the general case\label{A}}
\noindent We shall first consider the case $m=V=0$ where \eqref{eq54} reads
$$
  \tau_1(z) = 1 + \frac{1}{2r^2z} (z-1)^2 -\frac{z-1}{2r^2z} \sqrt{z^2-2\mu z+1}\,,
$$
with $\mu:=1-2r^2$. Here, the branch of the square root is chosen such that $\tau_1(z)=\mathcal O(|z|^{-1})$ as $z\to\infty$. 
Using 
\begin{equation}\label{eq:63}
  Z^{-1}\left\{ \frac{\sqrt{z^2-2\mu z+1}}{z} \right\} = P_{n-2}(\mu) - 2\mu P_{n-1}(\mu) + P_{n}(\mu) 
  = \frac{1}{n}[P_{n-2}(\mu)-\mu P_{n-1}(\mu)]\,,
\end{equation}
where $P_n$ denotes the Legendre polynomials (with the convention $P_{-1}=P_{-2}:=0$), we obtain
$$
  \tau_1^{(n)} = (1-\frac{1}{r^2})\delta_0^n + \frac{1}{2r^2}\delta_1^n 
  - \frac{1}{2r^2}\Big[P_{n+1}(\mu) - (2\mu+1) (P_{n}(\mu) - P_{n-1}(\mu)) - P_{n-2}(\mu)\Big],\;n\ge0\,.
$$
And this simplifies to $\tau_1^{(n)}=\delta_1^n$ for $r=1$.\\
\\
Next we shall discuss the general case with $m,\,V\in\R$. Here we have 
$$
  \tau_1(z) = 1 + \frac{\tilde c}{2z} -\frac{1}{2z}\sqrt{\tilde c}\sqrt{\tilde c+4z}\,,
$$
where $\tilde c(z):=z\,c(z)=\alpha z^2+\beta z+\gamma$ is a quadratic polynomial given by \eqref{M}.
Using again \eqref{eq:63}, both square root factors can be inverse Z-transformed. And the explicit formula for the coefficients $\tau_1^{(n)}$ would then involve a discrete convolution.\\
\\
But we shall proceed differently here and rather derive a recursion relation for $\tau_1^{(n)}$ (similar as in \S3 of \cite{EhAr01}).
A lengthy, but straightforward computation shows that $\hat\tau_1(z):=z\,\tau_1(z)$ satisfies the inhomogeneous differential equation
\begin{equation}\label{eq:64}
  \tilde c(\tilde c + 4z)\,\hat\tau_1'
  -[2\alpha^2z^3+3\alpha(\beta+2)z^2 + (\beta^2+4\beta+2\alpha\gamma)z+(\beta+2)\gamma]\,
  \hat\tau_1 = 2z(\gamma-\alpha z^2)\,.
\end{equation}
Since all coefficients in \eqref{eq:64} are polynomials we shall use the Laurent series of $\hat\tau_1$, i.e. $\hat\tau_1=\sum_{n=0}^\infty s^{(n)}z^{-n}$, with $\tau_1^{(n)}=s^{(n-1)}$. A comparison of the coefficients then yields:
$$
  s^{(0)}=\frac{1}{\alpha},\quad s^{(1)}=-\frac{\beta+2}{\alpha^2},\quad
  s^{(2)}=\frac{\beta^2+4\beta+5-\alpha\gamma}{\alpha^3}\,,
$$
and the exact recursion
$$
  (n+5)\alpha^2 s^{(n+3)}+ (2n+7)\alpha(\beta+2) s^{(n+2)}+  
  (n+2)(\beta^2+4\beta+2\alpha\gamma) s^{(n+1)}+ (2n+1)(\beta+2)\gamma s^{(n)}+
  (n-1)\gamma^2s^{(n-1)}=0\,,
$$
for $n\ge0$ with the convention $s^{(-1)}:=0$.\\
\\

\newpage
\section{Gauge-invariant introduction of the electromagnetic vector potential\label{B}}
%
%???- begin
Dirac equation Eq. \eqref{dirac-eq}  sofar has been written for a charged massive particle in a scalar potential.  For a general account of external electromagnetic fields and/or self-interaction both scalar and vector potential are needed.  A gauge invariant introduction of 
the electromagnetic vector potential $A(x,t)$ is executed by replacing  the complex 2-spinor $\mbox{\boldmath$\psi$}=\left(\begin{array}{c} u(x,t) \\v(x,t)\end{array}\right) $ in Eq. \eqref{dirac-eq} by 
$\mbox{\boldmath$\psi$}(x,t)\exp\{-i a(x,t)\} =\left(\begin{array}{c} u(x,t)\exp\{-i a(x,t)\} \nonumber \\v(x,t)\exp\{-i a(x,t)\}\nonumber \end{array}\right)$,   with\cite{peierls,graf}
$$
a(x,t):= \frac{q}{\hbar c}\int_{x_o}^{x} dy A(y,t)~.
$$
Position $x_o$ is arbitrary but constant.  Under this Peierls substitution the canonical momentum $p:= \frac{\hbar}{i}\frac{\partial}{\partial x}$ in  Eq. \eqref{dirac-eq} is replaced by the kinetic momentum $p - \frac{q}{c}A(x,t)= \frac{\hbar}{i}\frac{\partial}{\partial x}- \frac{q}{c}A(x,t)$. Under local gauge transformation $A(x,t)\rightarrow A(x,t)+ \frac{\partial}{\partial x} \Lambda(x,t)$, $\Phi(x,t)\rightarrow \Phi(x,t) - \frac{\partial}{c\partial t}\Lambda(x,t)$, the spinor acquires 
the phase factor $ \exp\{-i \frac{q}{\hbar c}\left(\Lambda(x,t)-\Lambda(x_o,t)\right)\}$, and the electromagnetic fields $E$ and $B$ remain invariant.\\
\\
Here it should be pointed out that in a (1+1)D model, orbital forces are confined to one spatial direction (i.e., $x$), while non-vanishing torque on the spin degree of freedom (in a Larmor term) arising from $A_x(x,t)=A(x,t)$ requires a non-vanishing $B$ field component in the plane orthogonal to $x$.  Note, however, that the two-component nature of an effective Dirac model may not arise from the spin degree of freedom.  Most notable example in 2+1D is graphene \cite{neto}.  This shows that the physical interpretation of the effective Dirac equation and the way an electromagnetic field couples to the system is determined by the underlying basic theory.\\ 
\\
We now discuss the consequences of the Peierls substitution on the leap-frog scheme Eqs. \eqref{eq:discretization-leap-frog-halfgrid1} and \eqref{eq:discretization-leap-frog-halfgrid}.  Again, we have $V(x,t)=-\Phi(x,t)$, and for any grid point $x_j,t_n$ we define $a^n_j:= a(x_j,t_n)$.  
The leap-frog scheme for non-zero vector potential is obtained by the substitution
\begin{align}
u^{n-1/2}_j &\rightarrow  {\hat u}^{n-1/2}_j := u^{n-1/2}_j \exp\{-i  a^{n-1/2}_j\}~, \nonumber \\ 
v^{n}_{j-1/2} &\rightarrow {\hat v}^{n}_{j-1/2} :=  v^{n}_{j-1/2} \exp\{-i  a^{n}_{j-1/2} \}~.
\label{eq:psub}
\end{align}
Likewise, the  stability analysis for zero vector potential detailed above can immediately be extended to the 
case of a non-vanishing vector potential by the  substitution Eq. \eqref{eq:psub} and noting that the vector potential $A(x,t)$ is real valued.  Hence, a strictly conserved functional ${\hat E}^n_r$  is identified by this substitution applied to the expression for  $E^n_r$ in Eq. \eqref{E} for arbitrary time $t$ and space $x$ dependent $m, V, A \in \mathbb{R}$.  It follows that the scheme remains stable for all $r=\Delta t /\Delta x\leq 1$.

Expressed in terms of spinor components $u$ and $v$ 	the scheme in presence of an external electromagnetic potential takes the form

\begin{align}
&f^+(a_j^{n+1/2},a_j^{n-1/2})\Bigg[ \frac{u_{j}^{n+1/2}-u_{j}^{n-1/2}}{\Delta t}+i\bigg(m_j^n-V_j^n -\frac{a_j^{n+1/2}-a_j^{n-1/2}}{\Delta t}\bigg)\frac{u_{j}^{n+1/2}+u_{j}^{n-1/2}}{2}\Bigg] \nonumber\\
&+ if^-(a_j^{n+1/2},a_j^{n-1/2})(m_j^n-V_j^n)\frac{u_{j}^{n+1/2}-u_{j}^{n-1/2}}{2} \nonumber \\
&+f^+(a_{j+1/2}^{n},a_{j-1/2}^{n})\Bigg[\frac{(D v^n)_j}{\Delta x}
-i \frac{a_{j+1/2}^{n}-a_{j -1/2}^{n}}{\Delta x}\frac{v_{j+1/2}^n+v_{j-1/2}^n}{2}\Bigg]=0\;,\label{eq:discretization-leap-frog-halfgrid2} 
\end{align}
\begin{align}
&f^+(a_{j-1/2}^{n+1},a_{j-1/2}^{n})\Bigg[\frac{v_{j-1/2}^{n+1}-v_{j-1/2}^{n}}{\Delta t}-i(m_{j-1/2}^{n+1/2}+V_{j-1/2}^{n+1/2}+\frac{a_{j-1/2}^{n+1}-a_{j-1/2}^{n}}{\Delta t})\frac{v_{j-1/2}^{n+1}+v_{j-1/2}^{n}}{2}\Bigg] \nonumber\\
&-i f^-(a_{j-1/2}^{n+1},a_{j-1/2}^{n})(m_{j-1/2}^{n+1/2}+V_{j-1/2}^{n+1/2})\frac{v_{j-1/2}^{n+1}-v_{j-1/2}^{n}}{2} \nonumber\\
&+f^+(a_{j}^{n+1/2},a_{j-1}^{n+1/2})\Bigg[\frac{(D u^{n+1/2})_{j-1/2}}{\Delta x}
-i\frac{a_{j}^{n+1/2}-a_{j-1}^{n+1/2}}{\Delta x} \frac{u_{j}^{n+1/2}+u_{j-1}^{n+1/2}}{2} \Bigg]=0 ~.
\label{eq:discretization-leap-frog-22}
\end{align}\\
\\
Here we have used the definition $f^\pm(a_1,a_2):=(e^{-i a_1}\pm e^{-i a_2})/2$.  As in the main text, we set $c=\hbar=1=-q$. For slowly varying vector potential (or within first order in $\Delta t$) one may approximate these equations by

\begin{align}
 \frac{u_{j}^{n+1/2}-u_{j}^{n-1/2}}{\Delta t}&+i(m_j^n-{\hat V}_j^n )\frac{u_{j}^{n+1/2}+u_{j}^{n-1/2}}{2}\nonumber\\
&+ \bigg[\frac{(D v^n)_j}{\Delta x}+i A_j^n\frac{v_{j+1/2}^n+v_{j-1/2}^n}{2}\bigg]=0\;,\label{eq:discretization-leap-frog-3}
\end{align}
%\end{document}
\begin{align}
\frac{v_{j-1/2}^{n+1}-v_{j-1/2}^{n}}{\Delta t}&-i(m_{j-1/2}^{n+1/2}+{\hat V}_{j-1/2}^{n+1/2})\frac{v_{j-1/2}^{n+1}+v_{j-1/2}^{n}}{2}
\nonumber \\
&+\bigg[\frac{(D u^{n+1/2})_{j-1/2}}{\Delta x}
+i A_{j-1/2}^{n+1/2} \frac{u_{j}^{n+1/2}+u_{j-1}^{n+1/2}}{2} \bigg]=0\;.
\label{eq:discretization-leap-frog-33}
\end{align}
Here we have used the following abbreviations on the two sub-lattices: ${\hat V}_j^n:=  V_j^n+ \frac{a_j^{n+1/2}-a_j^{n-1/2}}{\Delta t}$ and  $ {\hat V}_{j-1/2}^{n+1/2}=V_{j-1/2}^{n+1/2}+\frac{a_{j-1/2}^{n+1}-a_{j-1/2}^{n}}{\Delta t}$ denote the net scalar potential associated with the $E$ field after introduction of the vector potential,  and  
$ A_j^n=\frac{a_{j-1/2}^{n}-a_{j +1/2}^{n}}{\Delta x}$ and $A_{j-1/2}^{n+1/2}=\frac{a_{j-1}^{n+1/2}-a_{j}^{n+1/2}}{\Delta x}$ are the vector potential, as defined by symmetric spatial derivatives of $a(x,t)$.  Note that we use $q=-1$.  

The approximate scheme \eqref{eq:discretization-leap-frog-3} and \eqref{eq:discretization-leap-frog-33} may have been guessed directly by inspection of Eqs. \eqref{eq:discretization-leap-frog-halfgrid1} and \eqref{eq:discretization-leap-frog-halfgrid}.   Going the present way, however,  not only has given the way for  precise implementation of the vector potential into the latter but also has taken care of the stability analysis for this general case.  Finding an exactly conserved functional for the approximate scheme is complicated by the fact that the vector potential leads to additional coupling between the spinor components $u$ and $v$.
%
%???-end
\end{appendix}
\newpage
%
%
%
%-------BIBLIOGRAPHY-------
%
 
%
%
\end{document}